\documentclass[reprint,superscriptaddress,nofootinbib,amsmath,amssymb,aps]{revtex4-1}

\usepackage{graphicx}
\usepackage{dcolumn}
\usepackage{bm}
\usepackage{mathrsfs}
\usepackage{appendix}
\usepackage{tikz}
\usepackage{xcolor}
\usetikzlibrary{arrows.meta, positioning, shapes.geometric}
\usepackage[dvipsnames]{xcolor}

\def\apj{Astrophys. J.}

\def\apjs{Astrophys. J. Suppl. Series}
\def\prd{Phys. Rev. D}

\def\aap{Astron. Astrophys.}

\begin{document}

\preprint{}

\title{Neutrino Flux from Collisions of Red Giant Cores in Globular Clusters}

\author{Cecilia Romero Rodr{\'i}guez}
\affiliation{Higher Institute of Technologies and Applied Sciences, Havana University, Havana, Cuba}
\author{Pau Amaro Seoane}
\affiliation{Universitat Politècnica de València, Spain}
\affiliation{Max Planck Institute for Extraterrestrial Physics, Garching, Germany}
\affiliation{Higgs Centre for Theoretical Physics, Edinburgh, UK}

\date{\today}

\begin{abstract}
The coalescence of degenerate helium cores during red giant collisions—a process we term \textit{erythrohenosis}—introduces a novel class of transient astrophysical sources of high-energy neutrinos. Using stellar models generated with MESA and SPH simulations of the final inspiral phase, we develop a semi-analytical model to estimate the amount of hydrogen mixed into the cores, the energy release ($\approx 4.28 \times 10^{49}$ erg) that heats the remnant to $T_f \approx 5.3 \times 10^8$ K, the magnetic field amplification ($B \approx 1.77 \times 10^{10}$ G), and the resulting neutrino flux. The predicted TeV–PeV neutrino signal may contribute to the diffuse IceCube neutrino flux; we estimate that the horizon at which a single neutrino event is expected on average at $\gtrsim 100$ TeV is $\approx$ 55 kpc for individual merger events. Furthermore, we discuss the probability of a magnetized helium flash and assess the subsequent activation of the CNO cycle in the remnant core due to hydrogen mixing. In particular, neutrinos from the decay of $^{18}$F offer a direct observational test of the detonation, (as sources within $\sim$4.4 pc would produce a neutrino flux exceeding the solar CNO neutrino background). The simultaneous emission of high-energy hadronic neutrinos, gravitational waves, and---if the optical depth permits ($\tau_{\rm T}\sim 10^{14}$ in the core of the remnant) ---an electromagnetic signal would constitute a unique multimessenger signature of red giant core collisions, positioning erythrohenosis events as exotic yet potentially observable phenomena in dense stellar systems.
\end{abstract}

\maketitle

\section{Introduction}

Dense stellar environments, such as globular clusters and galactic nuclei, host exceptionally high stellar number densities, reaching $10^{6}$--$10^{8}$ pc$^{-3}$ \citep{AmaroSeoane_2023}. In these systems, close stellar encounters are frequent and can result in direct physical collisions, particularly involving evolved stars with large radii. Red Giants (RGs) are especially susceptible to such interactions due to their extended, loosely bound envelopes. Collisions between red giant stars (erythrohenosis) represent phenomena of extraordinary astrophysical interest; they are expected to trigger the rapid formation of a common envelope, followed by strong hydrodynamic dissipation and the inspiral of the degenerate cores, eventually forming an exotic binary system. Previous studies have primarily focused on the hydrodynamic evolution, mass ejection, and electromagnetic signatures of such events in globular clusters, where the relative velocities are low enough for the two giants to become on single object by the end of the collision \citep{seoane2026erythrohenosis}. However, the extreme thermodynamic conditions reached when the two cores collide at the end of the inspiral—including temperatures exceeding $10^8$ K, densities $\rho \sim 10^6$ g cm$^{-3}$, and the amplification of magnetic fields—suggest they may also be efficient sources of high-energy neutrino emission.

Neutrinos serve as privileged probes of these phenomena, carrying unaltered information from the deepest regions of the collision that remain inaccessible to electromagnetic observations. Due to their extremely small interaction cross sections and electrical neutrality, neutrinos escape the dense medium without absorption or scattering, preserving the spectral imprints of the extreme conditions generated during the impact. In this work, we focus on high-energy neutrino emission (TeV–PeV) arising from hadronic ($pp, p\alpha$) and photomesonic ($p\gamma$) interactions of protons accelerated in the magnetic field of the remnant. 

In addition to the high-energy signal, neutrino emission in erythrohenosis events is also driven by thermal processes—electron‑positron pair annihilation, plasmon decay, photoneutrino process, and nucleon‑nucleon bremsstrahlung—as well as weak nuclear reactions associated with ongoing nuclear burning and beta decays. While the detection of low‑energy ($\sim$MeV) neutrinos from extrasolar nuclear fusion remains extremely challenging—requiring either exceptionally luminous events or very nearby sources—we provide quantitative estimates of the expected fluxes from the $^{14}\mathrm{N}(\alpha,\gamma)^{18}\mathrm{F}$ decay chain (see e.g. \citep{Serenelli_2005, Irene}, \citep{Capelo_2023}) and from the subsequent activation of the CNO cycle. Furthermore, we investigate whether the strong magnetic field amplified during the merger could enhance the screening factor and thereby modify the nuclear reaction rate.

Neutrino emission from binary and collisional systems has been previously investigated (see, e.g., \cite{Cusinato_2022, Kimura_2018, Sekiguchi:2011zd, Christiansen:2005gw, Zhou_2023, foucart2025neutrinoscollidingneutronstars, 1998A&A...338..535R, Xiao:2016man}). We aim to build upon this background by searching for a detectable fingerprint from Earth using detectors such as KM3Net/ARCA \citep{Marinelli_2021}, Baikal \citep{Avrorin_2021, Baikal-GVD:2025rhg}, P-ONE \citep{resconi2021pacificoceanneutrinoexperiment}, Super-Kamiokande \citep{Super-Kamiokande:2002weg}, Hyper-Kamiokande \citep{protocollaboration2018hyperkamiokandedesignreport}, JUNO \citep{An_2016}, DUNE \citep{strait2016longbaselineneutrinofacilitylbnf}, Jinping \citep{Beacom2017Jinping}, and especially IceCube \citep{Aartsen_2017}, whose data we use to benchmark our model.

We develop a semi‑analytical framework that combines stellar evolution calculations (MESA) with smoothed‑particle hydrodynamics (SPH) simulations of the final inspiral. Using these tools, we characterize the thermodynamic conditions of the merging cores, the mixing of hydrogen into the degenerate material, and the amplification of the magnetic field via a turbulent dynamo once the cores collide. We then compute the expected high‑energy (TeV–PeV) neutrino flux from hadronic and photomesonic interactions of accelerated protons, as well as the lower‑energy ($\sim$MeV) signal from the $^{18}\mathrm{F}$ decay chain and the CNO cycle. The detectability of a single merger event and the contribution to the diffuse IceCube flux are assessed using the detector's effective area and the latest observational constraints on event rates in dense stellar systems. This work builds on \citep{seoane2026erythrohenosis} and is organized as follows. Section~\ref{sec:2} describes the MESA models and SPH simulations, Section~\ref{sec:3} quantifies the energy budget and magnetic field amplification and Section~\ref{sec:4} presents the hydrogen trapping model. The main results of this work can be found in Section~\ref{sec:5} where we calculate the high-energy neutrino production efficiency and detectability and in Section~\ref{sec:6} where we examine the helium flash, the magnetic screening of the $^{14}\mathrm{N}(\alpha,\gamma)^{18}\mathrm{F}$ reaction, and the post‑flash CNO cycle; Section \ref{sec:7} discusses the multimessenger implications and summarizes our conclusions.

\section{Simulating with MESA and SPH}\label{sec:2}

\begin{figure*}[t!]
    \centering
    \includegraphics[width=0.98\textwidth]{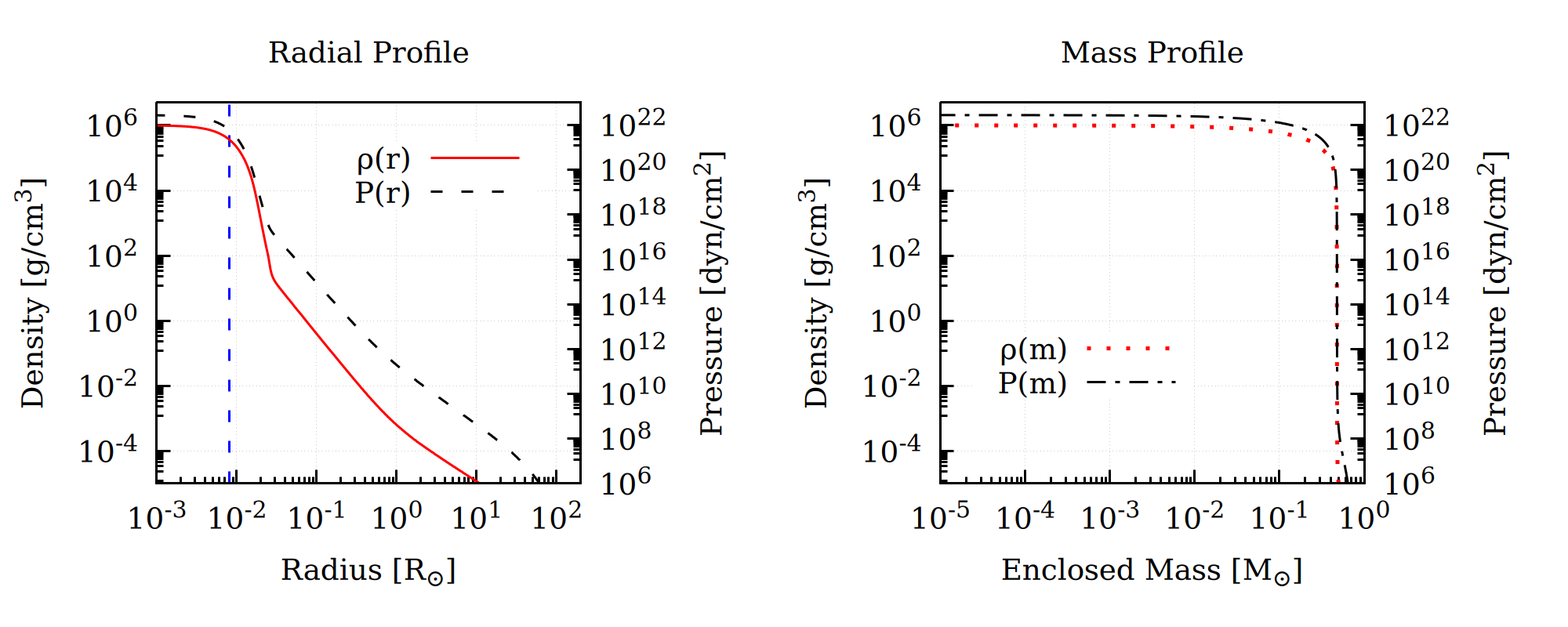}
    \caption{In the left panel we can see the Density (lower curve) and Pressure (upper curve) as a function of the stellar radius. The vertical line indicates the degenerate helium core. In the right panel we have the same quantities as a function of the enclosed mass, illustrating that the core contains nearly the entire mass of the star. We can appreciate  that these magnitudes remain constant within the nucleus, as expected for a red giant of this age.}
    \label{fig:profile40}
\end{figure*}

\begin{figure}[tbp]
    \centering
    \includegraphics[width=0.95\columnwidth]{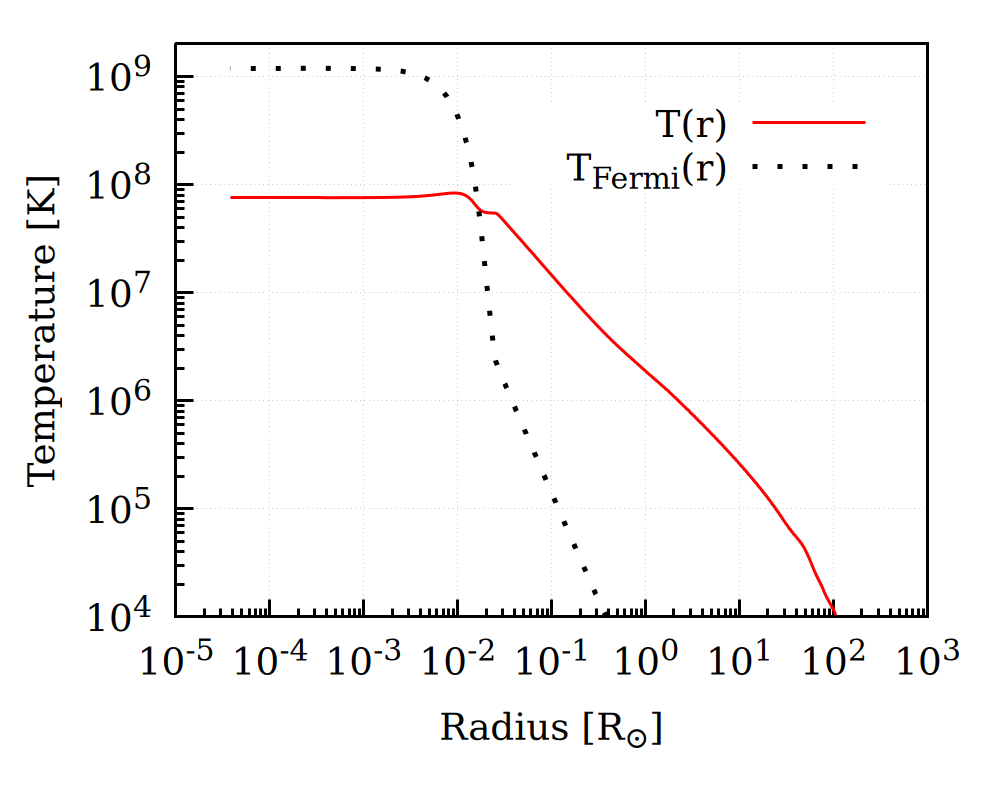}
    \caption{The solid red line represents temperature as a function of stellar radius. The temperature remains constant throughout the degenerate nucleus, reaching its maximum at the core boundary. A nearly isothermal region is visible between $2$--$3 \times 10^{-2}\text{ }R_\odot$, corresponding to the hydrogen-burning shell \citep{AmaroSeoane_2023}. The black-dotes curve display the Fermi temperature as a function of radius. These quantities converge at $\sim 2\times10^{-2} R_{\odot}$, approximately the size of the stellar core.}
    \label{fig:profile40T}
\end{figure}
We simulated a $1\text{ }M_\odot$ star with an initial metallicity of $2\times10^{-3}$, a typical value for globular cluster stars. The models were evolved using MESA \citep{MESA01, MESA02, MESA03, MESA04, MESA05, MESA06} until they reached masses of 0.85~$M_\odot$ and 0.69~$M_\odot$, respectively. These models then served as initial conditions for three-dimensional smoothed particle hydrodynamics (SPH) simulations to investigate the binary interaction between the degenerate cores. A detailed description of the stellar model generation and numerical setup is provided in \cite{seoane2026erythrohenosis}.

Our stellar models exhibit luminosities of $1.11\times10^{3}\text{ }L_\odot$ and $2.18\times10^{3}\text{ }L_\odot$, with effective temperatures of $3977\text{ K}$ and $3687\text{ K}$, respectively. The more massive, younger star has an age of $6.896\times10^{9}$~yr, while the older star, having experienced more significant mass loss during its evolution, has an age of $6.898\times10^{9}$~yr.

\begin{figure}[htbp]
    \centering
    \includegraphics[width=0.93\columnwidth]{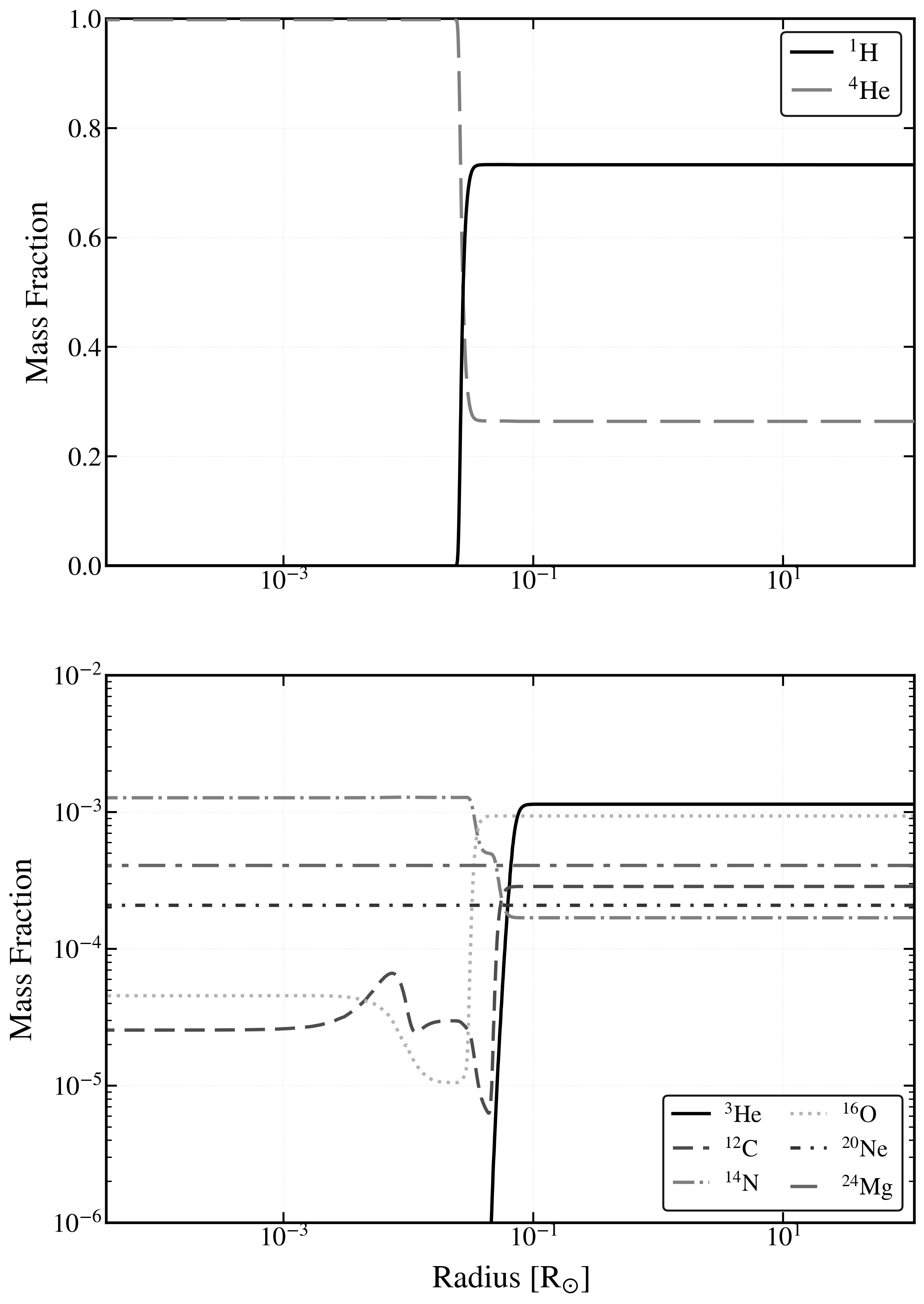}
    \caption{The upper panel displays the hydrogen and helium mass fractions as a function of stellar radius, clearly delineating the degenerate helium core; hydrogen becomes the predominant element upon reaching the burning shell. The lower panel shows the abundances of $^{3}\mathrm{He}$, $^{12}\mathrm{C}$, $^{14}\mathrm{N}$, $^{16}\mathrm{O}$, $^{20}\mathrm{Ne}$, and $^{24}\mathrm{Mg}$ from the MESA models. While the abundances of the latter two isotopes remain constant throughout the star, the others exhibit transitions at the hydrogen-burning shell, marking the boundary of the helium-degenerate region.}
    \label{fig:profile40abundancias}
\end{figure}

In this work, we focus primarily on the properties of the degenerate cores. Both cores extend to $\approx 8\times10^{-3}\text{ }R_\odot$, though the effective region involved in the binary interaction---the collision site---extends to $\sim 3\times10^{-2}\text{ }R_\odot$, where the hydrogen-burning shell is located. The pressure, density, and central temperature are nearly identical for both stars. Figures \ref{fig:profile40} and \ref{fig:profile40T} display the radial profiles of these quantities as functions of radius and enclosed mass for the 0.69~$M_\odot$ model, while Fig. \ref{fig:profile40abundancias} illustrates its chemical structure.

We first determine the width of the hydrogen-burning shell, which serves as the primary source of luminosity in a red giant by hosting thermonuclear reactions prior to the ignition of the triple-alpha process. The luminosity is given by:
\begin{equation}
    L = \int 4\pi r^2 \epsilon(r) \rho(r) dr,
\end{equation}
where $\epsilon(r)$ is the energy generation rate per unit mass per unit time obtained from the MESA output. By performing this integral numerically, we derive the characteristic radial extent of the hydrogen shell. For both models, this value is approximately $0.001\text{ }R_\odot$.

The Fermi energy before the collision is given by:
\begin{equation}
    E_{\mathrm{Fermi}} = \frac{\hbar^2}{2m_e}(3\pi^2n_e)^{2/3}.
\end{equation}
Assuming a helium mass fraction of $0.998$, the electron number density is $n_e \approx 3\times10^{29}\text{ cm}^{-3}$. The corresponding Fermi temperature is:
\begin{equation}
    T_{\mathrm{Fermi}} = \frac{E_{\mathrm{Fermi}}}{k_B} = 2.4 \times 10^9\text{ K}.
\end{equation}
Given that our averaged temperature is $\sim 8\times10^7\text{ K}$ for both stars, then $T \ll T_{\mathrm{Fermi}}$, confirming the presence of a fully degenerate (though non-relativistic) gas.

\begin{figure}[htbp]
    \centering
    \includegraphics[width=0.99\columnwidth]{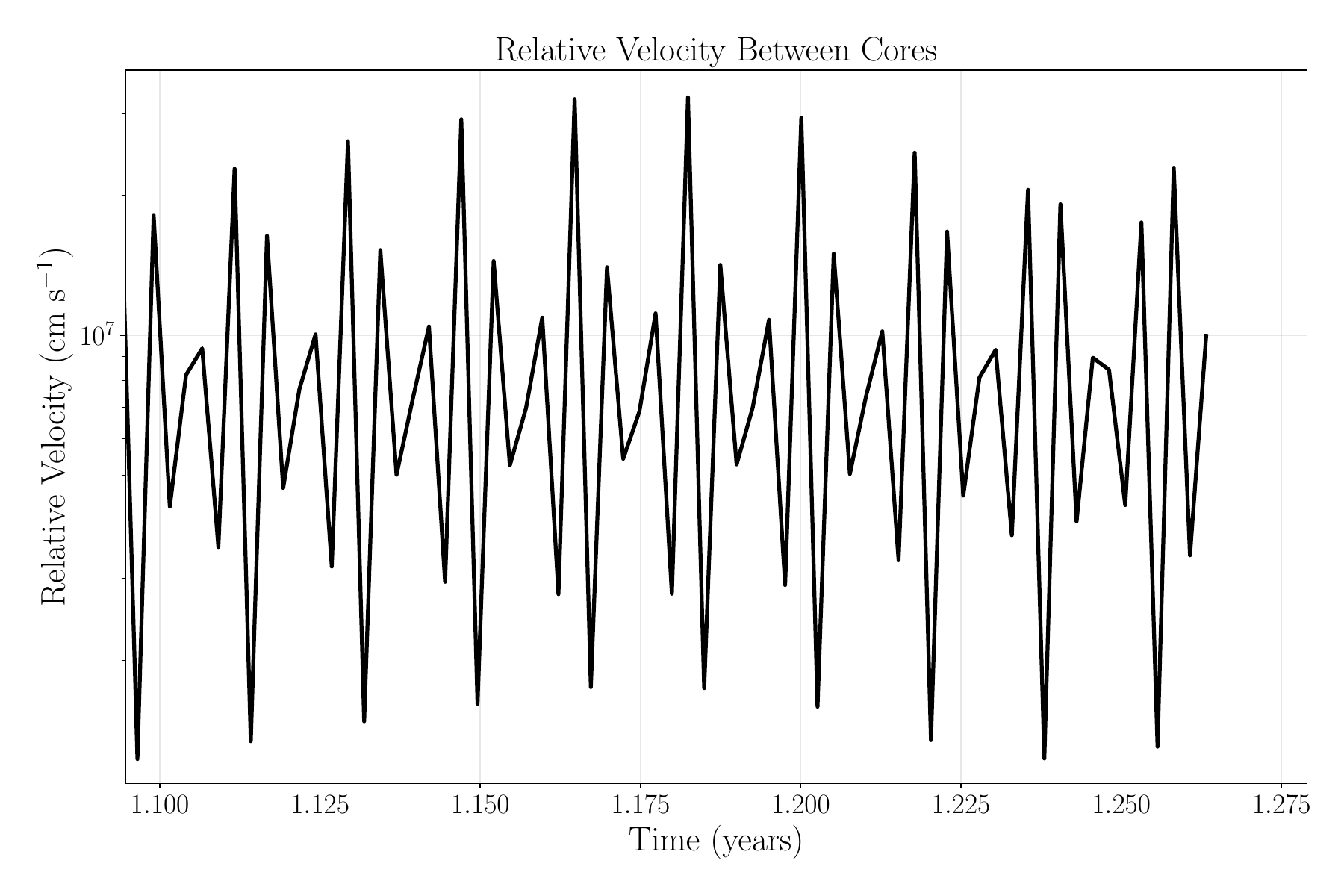}
    \caption{The relative velocity between the degenerate helium cores prior to merger. A non-zero eccentricity is evident; for this study, we adopt a characteristic value of $v_{\rm rel} = 100$~km~s$^{-1}$. This velocity determines the kinetic energy $E_{\rm coll}$ used in subsequent calculations.}
    \label{fig:mergers}
\end{figure}

Using the 3D SPH code, we simulated the collision and subsequent binary formation of the degenerate cores. Typical velocities for red giants encounters in globular clusters are on the order of $\sim 10$ km/s \citep{AmaroSeoane_2023}. However, we focus here not on the collisions of the stars themselves, but on the interaction of their degenerate nuclei. Figure~\ref{fig:mergers} illustrates the evolution of the relative velocity until first contact. One might expect the cores to accelerate further as they fall into each other’s gravitational potential well. However, the non-degenerate helium shell surrounding each core, with a density of $\sim 100\ \mathrm{g\ cm^{-3}}$, is sufficiently extended to counteract this effect. In this regime, both forces reach a dynamical equilibrium, preventing any significant net acceleration of the approaching cores prior to contact.

Regarding the initial stellar impact parameter, it influences the trajectory, it does not uniquely determine the parameter for the subsequent core collision. A detailed analysis of the eccentricity evolution is provided in \citep{seoane2026erythrohenosis}; however, as demonstrated below, a frontal collision serves as a robust approximation for the purposes of this work.

\section{Energy of the collision}\label{sec:3}
The collision of the two cores induces significant changes in the physical state of the gas, specifically raising the pressure, density, and temperature. To quantify these effects, we first consider the sound speed within the cores:
\begin{equation}
    c_s \sim \sqrt{\frac{\partial P}{\partial \rho}}.
\end{equation}
The relationship between pressure and density for a fully degenerate gas is $P = K\rho^{\gamma}$, where $\gamma = 5/3$ is the adiabatic index. Using the values from Fig.~\ref{fig:profile40} ($P = 2.5 \times 10^{22}$~dyn~cm$^{-2}$ and $\rho = 10^6$~g~cm$^{-3}$), we find $c_s \sim 2040$~km~s$^{-1}$. The corresponding Mach number is:
\begin{equation}
     \mathcal{M} = \frac{v_{\rm rel}}{c_s} \approx 0.05.
\end{equation}

Since the collision velocity is well below the sound speed of the degenerate core, the interaction does not produce a classical hydrodynamic shock. Instead, the core responds via quasi-elastic compression governed by electron degeneracy pressure. The fractional change in density scales as:
\begin{equation}\label{eq:lowdensityvariation}
    \frac{\Delta\rho}{\rho} \sim \mathcal{M}^2, \quad \mathcal{M} \ll 1.
\end{equation}
Consequently, both density and pressure remain nearly constant to a good approximation. Furthermore, since the ratio of the dynamical to signal-propagation timescales is proportional to the Mach number, information travels through the medium almost instantaneously relative to the crossing time. This implies that the final remnant quickly reaches a state of uniform $\rho$, $T$, and $P$.

As discussed in \cite{seoane2026erythrohenosis}, the energy released during the collision is governed by the center-of-mass kinetic energy:
\begin{equation}\label{eq:kinetic_energy_coll}
    E_{\rm kin} = \frac{1}{2} \mu v^2_{\rm rel},
\end{equation}
where $\mu = m_1 m_2 / M_{\rm m}$ is the reduced mass and $M_{\rm m} = m_1 + m_2$ is the total mass. Defining the mass asymmetry parameter $\eta = \mu / M_{\rm m}$, we obtain:
\begin{equation}\label{eq:kinetic_energy}
    E_{\rm kin} = \frac{1}{2} \eta M_{\rm m} v^2_{\rm rel}.
\end{equation}
This expression is valid for arbitrary mass ratios. The collision energy depends solely on the instantaneous relative speed at contact and is independent of orbital eccentricity, though the latter influences the value of $v_{\rm rel}$ (Fig.~\ref{fig:mergers}). Following first contact, the cores continue to approach one another due to gravitational attraction.

The internal energy of our models is given by:
\begin{equation}
    U_i = -\frac{3}{5} \frac{GM^2}{R},
\end{equation}
with an interaction term:
\begin{equation}
    U_d = -\frac{1}{2}\frac{GM^2}{R}.
\end{equation}
Given the relatively low collision velocity, we assume negligible mass loss. Since the density remains approximately constant, the volume $V$ is conserved, and it is reasonable to expect the formation of a single final object. Modeling the remnant as a sphere, we have:
\begin{align}
    V_{\rm c} &= 2 \cdot \left(\frac{4}{3}\pi R^3_{\rm c}\right), \\
    V_{\rm m} &= \frac{4}{3}\pi R^3_{\rm m},
\end{align}
which implies:
\begin{equation}
    R_{\rm m} = 2^{1/3}R_{\rm c}.
\end{equation}
The energy released in the collision is estimated as:
\begin{equation}\label{eq:deltaU}
    \Delta U = U_f - U_i \approx -4.28 \times 10^{49} \text{ erg},
\end{equation}
where $M_{\rm m} \approx 2M_{\rm c} = 0.78~M_\odot$ and $R_{\rm m} \approx 0.01~R_\odot$. This gravitational energy release is significantly higher than the kinetic energy ($\sim 10^{46}$~erg; see Eq.~\ref{eq:kinetic_energy} and \cite{seoane2026erythrohenosis}). This energy is partitioned into several channels: thermal energy, neutrino cooling, rotational energy, and magnetic field amplification.

\subsection{Magnetic field after merger}


Contact binaries typically exhibit significantly higher levels of magnetic activity compared to single stars of the same type \citep{vilhu1987binary}. Recent studies suggest that the magnetic field of a low-mass red giant near the hydrogen-burning shell ranges between 30--100~kG \citep{Li_2022}. Calculating the magnetic field of a planet or star is a highly complex task, even for numerical simulations; solving this for a stellar collision presents an even greater challenge.

\cite{ryu2025magneticfieldamplificationstellar} performed three-dimensional magnetohydrodynamic (MHD) simulations of collisions between low-mass main-sequence stars representative of old globular cluster populations. These simulations involved primary stars of $M_1 = 0.7$--$0.8\,M_\odot$ (near the turn-off at an age of $12\,\mathrm{Gyr}$) colliding with secondary stars with mass ratios $q = M_2/M_1 = 0.5$--$0.97$. All progenitor stars were evolved with \textsc{Mesa} assuming a low metallicity ($Z = 0.001$). Across a wide range of impact parameters ($b = 0.1$--$0.5$), the simulations revealed an amplification of magnetic energy by 8--10 orders of magnitude via small-scale dynamo action driven by turbulent and shearing flows. The resulting merger products exhibited saturated central magnetic field strengths of $B \sim 10^{7}$--$10^{8}\,\mathrm{G}$, largely independent of collision geometry or initial field strength, while surface fields reached $10$--$10^{4}\,\mathrm{G}$. The magnetic energy saturates near equipartition with the turbulent kinetic energy (see also \cite{planetary_nebulae_fromCEE}), indicating that magnetic back-reaction becomes dynamically important and naturally limits further amplification. This occurs because Lorentz forces modify the flow and halt growth. We can use this condition to estimate the upper limit of the magnetic field amplification:
\begin{equation}\label{eq:saturation_B}
    B^{2}/4\pi \sim \rho v^{2}.
\end{equation}

In the framework of small-scale turbulent dynamos, the classical analysis by \cite{Magnetic_field} emphasizes that magnetic amplification is primarily controlled by high-wavenumber components of the velocity field—specifically the small-scale eddies that dominate vorticity and line stretching. In this paradigm, the magnetic field saturates when its energy density becomes comparable to the kinetic energy of these small-scale motions. Under Kolmogorov’s hypotheses, this is set by the dissipation rate per unit mass, $\epsilon$, and the kinematic viscosity, $\nu$, rather than the large-scale velocity alone. This leads to a characteristic magnetic energy density of $\sim \rho \sqrt{\epsilon \nu}$, and Batchelor explicitly critiques simple equipartition arguments that equate magnetic energy with total kinetic energy across all scales.

However, for the present study, we do not require a detailed model of the turbulent cascade, which would necessitate specific models for the viscosity spectrum and kinematic viscosity—a task reserved for future work. Instead, we adopt a simpler equipartition estimate, assuming the magnetic field saturates at a level where its energy density is comparable to the characteristic kinetic energy density of the flow.

The field saturates when its energy density reaches a fraction of the turbulent kinetic energy density. Following criteria from small-scale dynamo simulations, we parameterize the saturation field as:
\begin{equation}
B_{\mathrm{sat}} = \sqrt{4\pi\rho}\,(\epsilon_B v_{\mathrm{coll}}),
\label{eq:B_sat_condition}
\end{equation}
where \(\epsilon_B\) is an efficiency factor typically in the range \(0.1\)--\(1.0\). For \(\epsilon_B = 0.5\), we obtain \(B_{\mathrm{sat}} \approx 1.77\times10^{10}\)~G and $E_{\rm B,sat} \approx 2\times 10^{46}$~erg.

To estimate the growth rate of the magnetic energy, we employ the MHD induction equation:
\begin{equation}\label{eq:magnetic_induction}
    \frac{\partial\mathbf{B}}{\partial{t}} = \nabla \times (\mathbf{v}\times \mathbf{B}) + \lambda \nabla^2\mathbf{B},
\end{equation}
where $\lambda = (\mu_0 \sigma)^{-1}$ is the magnetic diffusivity of the medium, $\mu_0$ is the vacuum magnetic permeability, and $\sigma$ is the electrical conductivity. The final term describes the average rate of conversion of magnetic energy into Joule heating.

A degenerate gas is an excellent conductor of electricity and heat; due to the extremely high densities, electrons are free and highly mobile. Consequently, we operate in the limit where $\sigma \rightarrow \infty$, or equivalently, the magnetic diffusivity $\lambda \rightarrow 0$. By neglecting the diffusion term in Eq.~(\ref{eq:magnetic_induction}) and expanding the right-hand side, we obtain the expression for the convective derivative (see \cite{Moffatt1978} and Appendix \ref{A}):
\begin{equation}\label{eq:derivada_convectiva}
    \frac{D\mathbf{B}}{Dt} = (\mathbf{B} \cdot \nabla)\mathbf{v},
\end{equation}
where we have assumed an incompressible fluid ($\nabla \cdot \mathbf{v} = 0$).

We adopt the center-of-mass frame without loss of generality. Since the flow remains subsonic ($\mathcal{M} \ll 1$) and the density varies only weakly during the interaction, the fluid can be treated as nearly incompressible. In this regime, the velocity field can be locally expanded around the collision point using a Taylor series:
\begin{equation}
\mathbf{v}(\mathbf{x}) = \mathbf{v}(\mathbf{0}) + (\nabla \mathbf{v})_{\mathbf{0}} \cdot \mathbf{x} + \mathcal{O}(x^2).
\end{equation}
The zeroth-order term vanishes by symmetry, leaving the linear term as the leading contribution to magnetic field amplification. 

A key result from \citet{ryu2025magneticfieldamplificationstellar} is that the saturation of the magnetic field is independent of the impact parameter. Therefore, for simplicity, we assume axial symmetry about the collision axis ($z$) and isotropy in the transverse plane. The most general linear velocity field consistent with incompressibility is:
\begin{equation}\label{eq:vel_sol_linear}
    v_x = ax, \qquad v_y = by, \qquad v_z = cz,
\end{equation}
where $a = b = \alpha$. The solenoidal condition $\nabla\cdot\mathbf{v}= a + b + c = 0$ implies $c = -2\alpha$. Thus, Eq.~(\ref{eq:vel_sol_linear}) becomes:
\begin{equation}
\label{eq:velocidades}
v_x = \alpha x, \qquad v_y = \alpha y, \qquad v_z = -2\alpha z,
\end{equation}
where $\alpha$ is the strain rate. The numerical coefficients are uniquely fixed by the requirement that compression along the collision axis be balanced by symmetric expansion in the transverse directions. Dimensional analysis suggests $\alpha \sim \tau^{-1}$, where $\tau$ is the characteristic timescale.

In the frontal, subsonic collision of the two degenerate helium cores, the velocity field in the interaction region is not limited by inertial braking or shock formation, but by the system's relaxation toward hydrodynamic equilibrium. In this regime, pressure forces efficiently redistribute momentum, enforcing a quasi-static adjustment of the flow. A convenient estimate for the duration of the dynamical phase is the sound-crossing time of the merged configuration \citep{Kippenhahn1994}:
\begin{equation}
    \tau_{\rm dyn} \sim \frac{R_{\rm m}}{c_{\rm s,m}}.
\end{equation}
This timescale corresponds to the fastest mechanical adjustment time of a self-gravitating body—the time required for pressure disturbances to propagate across the object—and represents the shortest macroscopic timescale on which the fluid can respond to external forcing.

Furthermore, Eq.~(\ref{eq:derivada_convectiva}) is interpreted in the Lagrangian sense along fluid trajectories. Over the dynamical timescale, the velocity gradient can be treated as approximately constant, allowing the convective derivative to be integrated as an ordinary time derivative:
\begin{equation}
\begin{aligned}
\frac{dB_x}{dt} &= \lambda_{\alpha}\frac{c_s}{R_{\rm m}} B_x,\\
\frac{dB_y}{dt} &= \lambda_{\alpha}\frac{c_s}{R_{\rm m}} B_y,\\
\frac{dB_z}{dt} &= -2\lambda_{\alpha}\frac{c_s}{R_{\rm m}} B_z,
\end{aligned}
\end{equation}
where we introduce the dimensionless parameter $\lambda_{\alpha}$. These equations admit the exact solutions:
\begin{equation}
\begin{aligned}
B_x(t) &= B_{x,0}e^{\lambda_{\alpha}\frac{c_s}{R_{\rm m}} t},\\
B_y(t) &= B_{y,0}e^{\lambda_{\alpha}\frac{c_s}{R_{\rm m}} t},\\
B_z(t) &= B_{z,0}e^{-2\lambda_{\alpha}\frac{c_s}{R_{\rm m}} t}.
\end{aligned}
\end{equation}
This indicates that magnetic field components perpendicular to the collision axis grow exponentially, while the longitudinal component decays. Assuming an initially isotropic field where $\langle B_{x,0}^2 \rangle = \langle B_{y,0}^2 \rangle = \langle B_{z,0}^2 \rangle = B_0^2/3$, the total field strength evolves as:
\begin{equation}\label{eq:magnetic_amplification}
B(\mathbf{r},t) = \sqrt{\frac{2}{3}} B_0 e^{\lambda_{\alpha}\frac{c_{\rm s,m}}{R_{\rm m}} t}.
\end{equation}

This solution predicts exponential amplification driven by local velocity gradients within the framework of ideal, kinematic MHD, where the field is passively advected and back-reaction is neglected. Consequently, this growth cannot persist indefinitely and must saturate at $B_{\rm sat}$.

\begin{figure}
    \centering
    \includegraphics[width=0.9\linewidth]{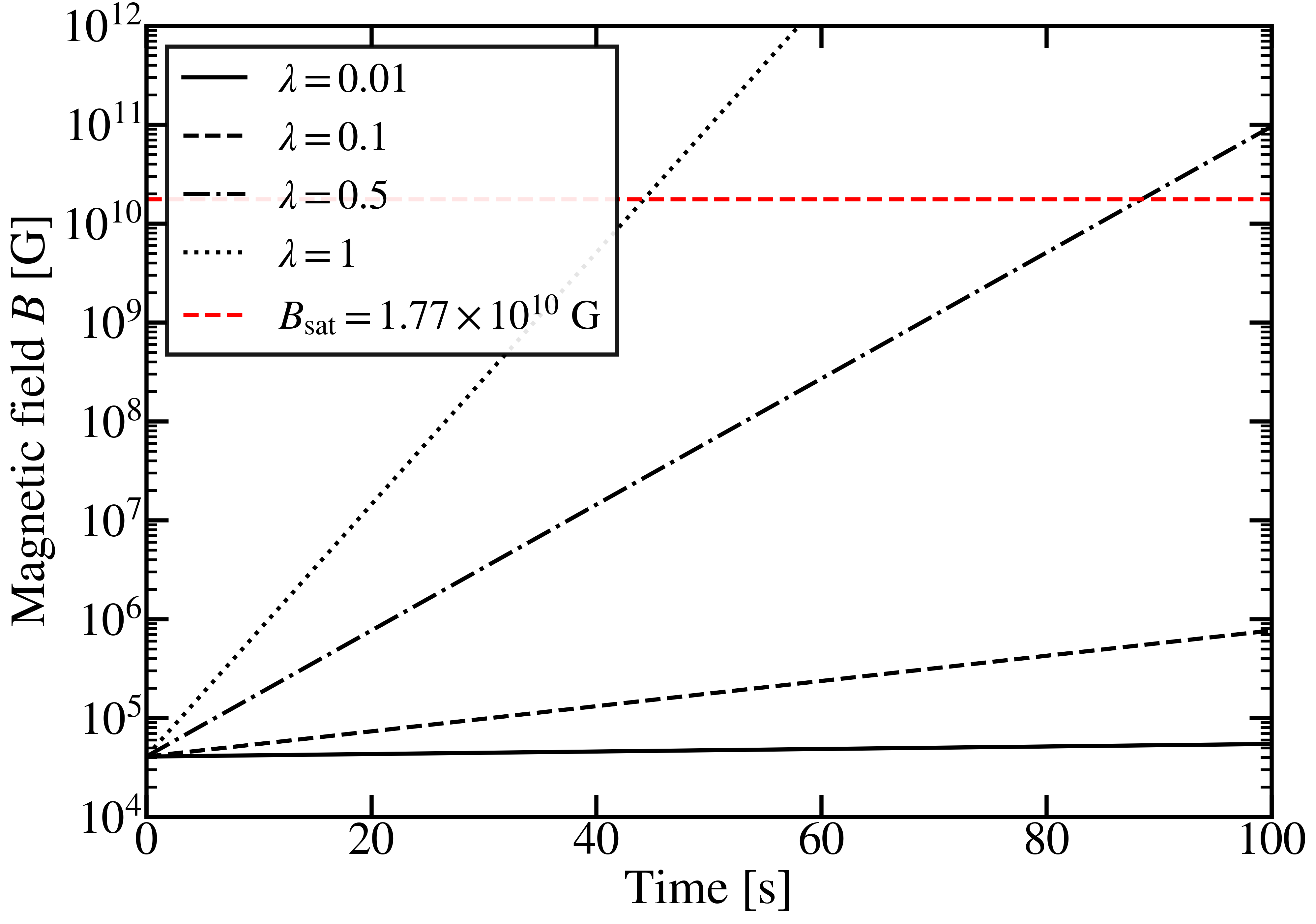}
    \caption{Evolution of the magnetic field for several values of the growth parameter $\lambda_{\alpha}$. For values $\leq 0.5$, saturation is not reached efficiently within the characteristic $\sim 100$~s timescale of the phenomenon.}
    \label{fig:magnetic_field_evolution}
\end{figure}

This behavior is consistent with numerical simulations of white dwarf collisions, which typically find maximum field strengths of $10^8$--$10^9$~G \citep{Ji_2013}. These values indicate that while significant amplification occurs, kinematic exponential growth is rapidly quenched by magnetic back-reaction and the finite duration of the interaction.

\subsection{Timescales}
We evaluate Eq.~(\ref{eq:magnetic_amplification}) using an initial field $B_0 \approx 50$~kG. With a sound speed in the degenerate core of $c_{\rm s,m} \approx 2040$~km~s$^{-1}$ and an effective collision radius of $R_{\rm m} \approx 0.01\,R_\odot$, the characteristic growth timescale is:
\begin{equation}
\tau_{\rm g} \approx 3.41\,\lambda_{\alpha}^{-1}\ \mathrm{s}.
\label{eq:tau_mag}
\end{equation}
For $\lambda_{\alpha} = 0.5$, $\tau_{\mathrm{g}} \approx 7$~s. This rapid growth is illustrated in Fig.~\ref{fig:magnetic_field_evolution}, where the field reaches saturation in $\sim 40$~s for $\lambda_{\alpha} = 1$. The saturation time $t_{\mathrm{sat}}$ is given by:
\begin{equation}
t_{\mathrm{sat}} = \tau_{\mathrm{g}} \ln\left( \frac{B_{\mathrm{sat}}}{B_0} \sqrt{\frac{3}{2}} \right).
\label{eq:t_sat}
\end{equation}

Estimating the characteristic merger timescale $\tau_{\rm m}$ for two degenerate cores is a non-trivial task. \cite{2012ApJShen} demonstrate that the evolution of white dwarf merger remnants is dominated by the transport of angular momentum and entropy. They estimate that magnetic stresses lead to viscous evolution over $10^4$--$10^8$~s for masses between $0.6$--$0.9~M_\odot$. Our scenario is qualitatively similar, though involving lower-mass cores; thus, we adopt a conservative estimate of $\tau_{\rm m} \sim 10^3$~s.

Comparing this with Fig.~\ref{fig:magnetic_field_evolution}, we find that magnetic saturation occurs well before the merger completes, provided $\lambda_{\alpha}$ is not excessively small. These cases are of primary interest: the field reaches its maximum strength while the cores are still merging. The exact value of $\lambda_{\alpha}$ is less critical than whether saturation is achieved within the merger window. This is particularly relevant because a helium flash may occur if the temperature rises sufficiently—an outcome we demonstrate is highly probable. While such an event would significantly influence the magnetic field evolution and necessitates detailed simulations, for the purposes of this work, we restrict our analysis to the phase preceding helium detonation.

\subsection{Thermal Neutrino Losses}
A fraction of the gravitational energy released during the merger is lost via thermal neutrino emission. In addition to neutrinos produced via fusion or nuclear decay, thermal processes in the stellar plasma contribute to the cooling. While these loss rates are well-characterized for stellar interiors, their impact on the coalescence timescale must be quantified to validate the temperature evolution derived later.

In the hot, dense environment of the merged core, several production mechanisms operate: pair annihilation ($e^+e^-\to \nu\bar{\nu}$), the photoneutrino process ($\gamma e^-\to e^-\nu\bar{\nu}$), the plasma process (plasmon decay into $\nu\bar{\nu}$), and bremsstrahlung ($e^- + (Z) \to e^- + (Z) + \nu\bar{\nu}$). For strongly degenerate electrons ($T \ll T_{\mathrm{F}}$) at temperatures of $\sim 10^8$~K, the plasma neutrino process is the dominant cooling channel. The energy-loss rate per unit volume for the plasma neutrino process is given by \cite{1996ApJSItoh}:

\begin{equation}\label{eq:Qplasma}
Q_{\mathrm{plasma}} = \bigl( C_V^2 + n C_V'^{,2} \bigr) Q_{\rm V},
\end{equation}
where $C_{\rm V} = \frac{1}{2} + 2\sin^2\theta_W$ and $C_{\rm V}' = -\frac{1}{2} + 2\sin^2\theta_W$ and $n = 2$. Equations (23)-(27) of \cite{1994ApJHaft} will be used for the calculation of $Q_{\rm V}$:
\begin{equation}
Q_{\rm V}= 3.00\times10^{21}\lambda^{9}\gamma^{6}e^{-\gamma}(f_T + f_L)f_{xy}.
\end{equation}

The instantaneous neutrino luminosity is then:
\begin{equation}
\dot{Q}_{\nu} = Q_{\mathrm{plasma}}\,V \approx 1.8\times10^{34}\text{ erg s}^{-1},
\end{equation}
where we used $Q_{\mathrm{plasma}} \approx 1.3\times10^{7}\ \mathrm{erg\ cm}^{-3}\mathrm{s}^{-1}$.
Integrating over the merger duration, the total energy lost to thermal neutrinos is $E_{\nu,\mathrm{therm}} = \dot{Q}_{\nu}\,\tau_{\mathrm{m}} \approx 8\times10^{37}$~erg. This is negligible compared to the gravitational energy release ($\Delta U$); thus, thermal neutrino emission removes an insignificant fraction of the available energy on the dynamical timescale.

\subsection{Rotational Energy Loss at High Impact Parameters}
The coalescence of two cores with a non-zero impact parameter $b$ results in the retention of orbital angular momentum, which is converted into rotational energy. For a large impact parameter $b = 0.9\,R_{\rm c}$, the orbital angular momentum at first contact is:
\begin{equation}
L_{\mathrm{orb}} = \mu v_{\mathrm{rel}} b \approx 1.83\times10^{48}\text{ g cm}^2\text{ s}^{-1},
\end{equation}
with a reduced mass $\mu \approx 0.2~M_\odot$. Assuming the final remnant relaxes to a spherical shape with radius $R_{\rm m}$, its moment of inertia is $I_{\rm m} = \frac{2}{5} M_{\mathrm{m}} R_{\rm m}^2 \approx 2.94\times10^{50}\text{ g cm}^2$. The resulting rotational energy is:
\begin{equation}
E_{\mathrm{rot}} = \frac{L_{\mathrm{orb}}^2}{2 I_{\rm m}} = 5.70\times10^{45}\text{ erg}.
\end{equation}

This accounts for only $\sim 0.013\%$ of the gravitational energy release. Even with an elongated geometry, which would increase the moment of inertia, the rotational energy remains negligible. Consequently, rotational energy does not significantly alter the global energy budget or the subsequent temperature evolution. This reinforces our conclusion that the remnant inevitably reaches temperatures well above the helium ignition threshold.

\subsection{Thermal Energy Evolution}
The gravitational energy released during the coalescence heats the plasma at an approximately constant rate, \(\dot{Q}_{\mathrm{grav}} = \Delta U_{\rm th} / \tau_{\mathrm{m}}\), over the merger timescale \(\tau_{\mathrm{m}}\). The available thermal energy is given by:
\begin{equation}
\Delta U_{\mathrm{th}} = \Delta U - \left( E_{\rm B,sat} + E_{\rm \nu, th} + E_{\mathrm{rot}} \right).
\end{equation}
Using the values derived in the preceding sections, we obtain $\Delta U_{\rm th} \approx 4.26 \times 10^{49}$~erg. Assuming that this energy is instantaneously thermalized and that the heat capacity remains constant, the temperature increases linearly with time:
\begin{equation}
T(t) = T_i + \frac{T_f - T_i}{\tau_{\mathrm{m}}}\, t, \label{eq:T_linear}
\end{equation}
where \(T_i\) is the initial temperature of the degenerate cores and \(T_f\) is the final mean temperature reached at the end of the merger. For a helium-dominated environment, we find:
\begin{equation}
U_{\mathrm{th}}(T) = \frac{3}{2} N_i k_B T
                 + \frac{\pi^2}{4} N_e \frac{(k_B T)^2}{E_F},
\end{equation}
where $N_i$ is the number of ions and $N_e = 2N_i$ for fully ionized helium. Defining $A \equiv 3/2 N_i k_B$ and $B \equiv \frac{\pi^2}{4} N_e \frac{k_B^2}{E_F}$, the energy required to raise the temperature from $T_i$ to $T_f$ is the quadratic equation
\begin{equation}
\Delta U_{\mathrm{th}}(T_f) = A(T_f - T_i) + B(T_f^2 - T_i^2),
\end{equation}
which yields:
\begin{equation}
T_f \approx 5.3\times10^8\,\mathrm{K}.
\end{equation}

This final temperature \(T_f\) exceeds the threshold for helium ignition (\(T_{\mathrm{ign}} \approx 2 \times 10^{8}\,\mathrm{K}\); \citep{Kippenhahn1994}) by a significant margin.

The time \(t_{\mathrm{ign}}\) at which the temperature first reaches the ignition threshold follows from Eq.~(\ref{eq:T_linear}):
\begin{equation}
t_{\mathrm{ign}} = \tau_{\mathrm{m}}\,\frac{T_{\mathrm{ign}} - T_i}{T_f - T_i} \approx 196\,\mathrm{s}. \label{eq:t_ign}
\end{equation}
Thus, helium ignites approximately \(196\,\mathrm{s}\) after first contact of the degenerate cores, well before the merger process is completed.

This simultaneity has significant implications: the helium flash ignites in a medium already supporting a magnetic field of \(\sim 10^{10}\,\mathrm{G}\). While the magnetic pressure remains negligible compared to the gas pressure, the field may influence subsequent convection, turbulence, and neutrino emission channels. In particular, such a strong field ensures that high-energy hadronic processes operate in a magnetized environment, potentially affecting the cooling rates of secondary pions and muons.


\section{Hydrogen penetration in the degenerate cores during the merger}\label{sec:4}

A peculiar phenomenon arises from the collision: while a red giant at this evolutionary stage typically has no hydrogen remaining in its core, the approaching cores eventually bring their hydrogen-burning shells into contact. These shells begin to compress, leading to two possible outcomes: the hydrogen may be transversely expelled, or it may remain confined and mix into the remnant. We develop a model to estimate the mass of hydrogen trapped between the two degenerate helium cores during their final approach and subsequent merger.

We define the $z$-axis as the collision axis and $h(r,t)$ as the instantaneous separation between the surfaces of the two degenerate helium nuclei. The time $t$ is initialized when the non-degenerate helium and hydrogen-fusing shells form a common envelope. The temperature of this common shell remains approximately consistent with the individual stellar models (see Fig.~\ref{fig:profile40T}), $T_H \approx 5\times10^7\,\mathrm{K}$. Based on numerical simulations of the final inspiral phase, we adopt a relative approach velocity of $v_{\rm rel} = 100$~km~s$^{-1}$.

The global geometry involves two spherical helium cores of radius $R_{\rm c}$, each surrounded by a shell of non-degenerate helium $\approx 0.02\,R_\odot$ wide and a thinner hydrogen-rich layer of $\approx 0.001\,R_\odot$. However, the hydrodynamics of hydrogen trapping occur only very close to the symmetry axis immediately prior to contact. In this regime, we use a local approximation where global parameters like the impact parameter can be neglected. Due to the spherical curvature, the separation increases with the distance $r$ from the axis. For $r \ll R_c$ and $h(t) \ll R_c$, we apply a parabolic approximation:
\begin{equation}\label{eq:h_of_r}
h(r,t) = h(t) + \frac{r^2}{R_c}.
\end{equation}

Only gas located near the symmetry axis experiences effective compression; gas at larger radii encounters a rapidly widening channel and escapes laterally. We define the characteristic transverse scale for geometric confinement as the region where the increase in separation remains comparable to the axial gap:
\begin{equation}
R_{\rm eff}(h) \equiv \delta\sqrt{R_{\rm c}\,h(t)},
\label{eq:Reff_def}
\end{equation}
where we adopt $\delta = 0.01$. Only hydrogen within $r < R_{\rm eff}$ participates in the transient "slab-like" compression. Assuming a constant relative velocity, the gap evolves as $h(t) = h_0 - v_{\mathrm{rel}}\, t$ for $0 \le t \le h_0/v_{\mathrm{rel}}$, where $h_0$ is the separation at first contact of the hydrogen shells.

Let $\rho_H(t)$ be the uniform density of hydrogen within the slab. We approximate the volume of the compressed region as a cylinder with radius $R_{\mathrm{eff}}$ and height $h$:
\begin{equation}
V = \pi R_{\mathrm{eff}}^2 h = \pi \delta^2 R_c h^2.
\end{equation}
Hydrogen leaves the slab through the lateral surface at velocity $v_\perp$ and is absorbed by the degenerate cores through the circular bases at velocity $v_{\mathrm{in}}$. Integrating the mass conservation equation $\frac{\partial \rho}{\partial t} + \nabla \cdot (\rho \mathbf{v}) = 0$ over the volume $V(t)$, we obtain:
\begin{equation}
\frac{d}{dt}(\rho V) = -\rho v_\perp \left( 2\pi \delta \sqrt{R_c} h^{3/2} \right) - \rho v_{\mathrm{in}} \left( 2\pi \delta^2 R_c h \right).
\end{equation}

Using $d/dt = -v_{\mathrm{rel}} d/dh$ and expanding the derivative, the density evolution is governed by:
\begin{equation}
\frac{1}{\rho_{\rm H}}\frac{d\rho_{\rm H}}{dh} = \frac{2v_\perp}{\delta \sqrt{R_c} v_{\mathrm{rel}}} h^{-1/2} + \frac{2v_{\mathrm{in}}}{v_{\mathrm{rel}}} h^{-1} - \frac{2}{h}.
\end{equation}
Integrating from $h_0$ to $h$ yields the density profile:
\begin{equation}
\rho_{\rm H}(h) = \rho_{\rm H,0} \left( \frac{h}{h_0} \right)^{\frac{2v_{\mathrm{in}}}{v_{\mathrm{rel}}} - 2} \exp \left[ \frac{4v_\perp}{\delta \sqrt{R_c} v_{\mathrm{rel}}} \left( \sqrt{h} - \sqrt{h_0} \right) \right]. \label{eq:density}
\end{equation}
With an initial hydrogen mass fraction $X = 0.1$, we take $\rho_{\rm H,0} \approx 10\,\mathrm{g\,cm^{-3}}$. The mass absorbed by the cores is then:
\begin{equation}
M_{\mathrm{H,abs}}(h) = \frac{2\pi \delta^2 R_c v_{\mathrm{in}}}{v_{\mathrm{rel}}} \int_{h}^{h_0} \rho_{\rm H}(h') \, h' \, dh'.
\end{equation}
Substituting Eq.~(\ref{eq:density}), we find:
\begin{align}
M_{\mathrm{H,abs}}(h) &= \frac{2\pi \delta^2 R_c v_{\mathrm{in}} \rho_{\rm H,0}}{v_{\mathrm{rel}} \, h_0^{\frac{2v_{\mathrm{in}}}{v_{\mathrm{rel}}}-2}} \int_{h}^{h_0} (h')^{\frac{2v_{\mathrm{in}}}{v_{\mathrm{rel}}}-1} \times\nonumber \\
&\exp \left[ \frac{4v_\perp}{\delta \sqrt{R_c} v_{\mathrm{rel}}} \left( \sqrt{h'} - \sqrt{h_0} \right) \right] dh'.
\end{align}


The lateral escape of hydrogen from the compressed region is driven by pressure gradients. To estimate the outflow velocity \(v_\perp\), we consider the radial component of the Euler equation, assuming axisymmetry and neglecting viscosity:
\begin{equation}
\frac{\partial v_r}{\partial t} + v_r \frac{\partial v_r}{\partial r} = -\frac{1}{\rho}\frac{\partial P}{\partial r}.
\end{equation}
During the compression phase, the characteristic time for lateral adjustment is \(\tau_{\mathrm{lat}} \sim R_{\mathrm{eff}}/c_{\rm s,s}\), while the compression time is \(\tau_{\mathrm{comp}} \sim h/v_{\mathrm{rel}}\). Given that \(R_{\mathrm{eff}}/c_{\rm s,s} \ll h/v_{\mathrm{rel}}\), the lateral flow reaches a quasi-steady state on the compression timescale, allowing us to neglect the partial time derivative \(\partial v_r/\partial t\).

Using the effective interaction radius \(R_{\mathrm{eff}}\) as the radial length scale and defining \(v_\perp \equiv v_r\), we approximate:
\begin{equation}\label{eq:eulerr}
v_r \frac{\partial v_r}{\partial r} \sim \frac{v_\perp^2}{2R_{\mathrm{eff}}}. 
\end{equation}
Assuming the pressure drops from a central value \(P\) (at \(r=0\)) to \(P_{\mathrm{eff}}\) at the boundary \(r = R_{\mathrm{eff}}\), the pressure gradient scales as:
\begin{equation}
\frac{\partial P}{\partial r} \sim \frac{P - P_{\mathrm{eff}}}{R_{\mathrm{eff}}}. 
\end{equation}
Combining these with the ideal gas equation of state \(P = \rho c_{\rm s,s}^2 / \gamma\), we find:
\begin{equation}\label{eq:presiones}
v_\perp^2 \sim 2c_{\rm s,s}^2 \left(1 - \frac{P_{\mathrm{eff}}}{P}\right). 
\end{equation}

Defining the dimensionless parameter \(\xi \equiv P_{\mathrm{eff}}/P\) (where \(0 < \xi < 1\)), the outflow velocity becomes:
\begin{equation}
v_\perp \sim \sqrt{2(1-\xi)}\; c_{\rm s,s}.
\end{equation}
The value of \(\xi\) is dictated by the confinement geometry. In our scenario, the curved surfaces of the degenerate cores and the surrounding stellar material prevent complete pressure release. We adopt a moderate pressure drop of \(\xi = 0.95\), yielding \(v_\perp = 0.32\,c_{\rm s,s}\). 

Under the conditions in the shell surrounding the degenerate helium core, the gas acts as a fully ionized plasma with a mean molecular weight \(\mu \simeq 1.14\). The shell pressure is:
\begin{equation}
P = \frac{\rho_H}{\mu m_p} k_B T \simeq 3.6 \times 10^{17}\ \mathrm{dyn\ cm^{-2}},
\end{equation}
leading to a sound speed of:
\begin{equation}
c_{\rm s,s} = \sqrt{\gamma \frac{P}{\rho}} \sim 776\ \mathrm{km\ s^{-1}}.
\end{equation}

To estimate the inward absorption velocity \(v_{\mathrm{in}}\) into the degenerate cores, we adopt a phenomenological scaling:
\begin{equation}
\frac{v_{\mathrm{in}}}{v_{\mathrm{rel}}} = \eta \, \frac{c_{\rm s,s}}{v_{\mathrm{rel}}} \left( \frac{P_{\mathrm{shell}}}{P_{\mathrm{core}}} \right)^{1/2},
\end{equation}
where \(\eta = 0.32\) maintains consistency with the treatment of \(v_{\perp}\). Using the fiducial values \(P_{\mathrm{shell}} = 3.6 \times 10^{17}\ \mathrm{dyn\ cm^{-2}}\) and \(P_{\mathrm{core}} = 2.5 \times 10^{22}\ \mathrm{dyn\ cm^{-2}}\), we find \(v_{\mathrm{in}}/v_{\mathrm{rel}} \approx 0.0094\), or \(v_{\mathrm{in}} \approx 0.94\ \mathrm{km\ s^{-1}}\). This absorption rate results in a trapped hydrogen mass of approximately $4 \times 10^{-14}$ M$_\odot$. Assuming this mass mixes uniformly throughout the remnant volume \(V_{\rm m}\), the resulting proton density is:
\begin{equation}
    n_p = \frac{M_{\rm abs,H}}{V_{\rm m} m_{\rm H}} \approx 1.35 \cdot 10^{17}\ \mathrm{cm^{-3}}.
\end{equation}

While this geometric model provides a conservative lower limit, the actual hydrogen mass is likely higher due to hydrodynamic instabilities. Specifically, the Kelvin-Helmholtz (KH) instability and the Rayleigh-Taylor (RT) instability.
KH instability arises from the velocity shear across the interface between the two cores and the surrounding envelope. As the cores approach, the relative motion generates strong tangential velocity gradients, causing the interface to roll up into characteristic vortices. These vortices entrain hydrogen-rich material from the envelope and mix it downward into the helium core, effectively eroding the compositional boundary layer. 
On the other hand, RT inestability is driven by the deceleration of the denser helium core against the lower density envelope.
These processes would disrupt the laminar boundary layers and entrain significantly more hydrogen-rich material into the degenerate core. Nevertheless, our model demonstrates that even with efficient leakage, a non-negligible amount of hydrogen is trapped, providing the necessary target material for hadronic neutrino production.

\section{Neutrino production}\label{sec:5}

This section addresses the physical mechanisms that yield a detectable neutrino flux. In high-energy astrophysical environments, three primary channels are efficient at producing neutrinos: inelastic proton--proton (\(pp\)) interactions, inelastic proton--alpha (\(p\alpha\)) interactions, and the photomeson production process (\(p\gamma\)). The \(pp\) channel proceeds via the production and decay of secondary pions: \(p + p \rightarrow \pi^{\pm} + X\), with charged pions decaying as \(\pi^+ \rightarrow \mu^+ + \nu_\mu\) and subsequently \(\mu^+ \rightarrow e^+ + \bar{\nu}_\mu + \nu_e\). The $p\alpha$ channel is particularly relevant in the helium-rich environment of the degenerate core and proceeds via $p+\alpha \rightarrow \pi^{\pm}+X$. The \(p\gamma\) channel, dominated by the \(\Delta\)-resonance (\(p + \gamma \rightarrow \Delta^+ \rightarrow \pi^0 + p\) or \(\pi^+ + n\)), also populates the muon and electron neutrino families. Other interaction channels contribute only marginally and, in practice, suppress rather than enhance the final neutrino spectrum.

The origin of the high-energy neutrino flux (TeV -- PeV) must be carefully considered, as they don't escape as easily as low-energy neutrinos ($\sim$MeV) from the core of the remnant. First, they are expected to originate from charged pion decays produced via hadronic collisions in the post-merger core, and they can experience severe synchrotron losses. In the global saturated field $B_{sat} \approx 1.77 \times 10^{10}$G, the synchrotron cooling time for a pion falls orders of magnitude below the pion decay time $t_{\rm \pi,dec} = \gamma_\pi \tau_\pi$, where $\gamma_\pi = E_\pi/(m_\pi c^2)$. Consequently, pion decay is completely suppressed by synchrotron cooling in this field. In addition, at TeV–PeV energies, neutrinos produced in the core may undergo significant attenuation, as their mean free path decreases due to the approximately linear increase of the neutrino–nucleon cross section with energy \citep{Valera2022}. The condition for efficient neutrino production, $t_{\pi,\rm syn}(B, E_\pi) > t_{\pi,\rm dec}$, therefore requires a substantially weaker magnetic field. We return to this condition below, after establishing the mechanism responsible for proton acceleration.

In the core of the remnant, the Thomson optical depth can be estimated as $\tau_{\rm T}=n_e \sigma_{\rm T} L$,
which yields $\tau_{\rm T}\sim10^{14}$ for the characteristic densities and length scales considered in this work. An optically thick medium might imply that any shocks developing within the dense core are likely radiation-mediated. In such a regime, standard diffusive shock acceleration may indeed be suppressed because photon diffusion smooths the velocity discontinuity.

However, efficient particle acceleration does not necessarily require collisionless shocks. The violent merger of two magnetized degenerate cores naturally generates strong magnetic compression, shear flows, turbulence, and complex topologies, as discussed in \ref{sec:3}, that drive magnetic field lines of opposite polarities together. These conditions favor the formation of extended current sheets.

Besides, given the extremely large Lundquist numbers ($S = \frac{L v_A}{\lambda} \approx 10^{15}$) expected in such dense and highly conducting plasma, the current sheets become unstable to fast plasmoid-mediated magnetic reconnection. Indeed, in the work of \cite{HuangBhattacharjee2010}, it is shown that current sheets at \(S > 10^4 \) cannot remain in a slow, stable Sweet-Parker state. Nevertheless, to rigorously prove the existence of such reconnection in a 3D context, MHD simulations will ultimately be required. In this work, however, we restrict ourselves to physically motivated arguments.

On the other hand, \cite{Uzdensky2011} emphasized that the extreme Thomson optical depth would suppress microscopic flows via photon drag or radiation viscosity. We are going to demonstrate that the two processes operate on entirely different length scales in our scenario. The mean free path of a photon in the post-merger core is $  \lambda_\text{ph} = \frac{1}{n_e \sigma_T}  
    \approx 5\times10^{-6}~\text{cm},$
whereas the characteristic scale of the reconnection layer is the ion inertial length,
\begin{equation}
d_i \equiv \frac{c}{\omega_{pi}}, \qquad
\omega_{pi} = \left(\frac{4\pi n_i Z^2 e^2}{m_i}\right)^{1/2},
\end{equation}
where $n_i$ is the ion number density, $Z$ the ion charge, and $m_i$ the ion mass. For the post-merger helium plasma, taking $n_i \approx 1.5\times10^{29}\,\mathrm{cm^{-3}}$, $Z=2$, and $m_i=m_{\mathrm{He}}$, we obtain $d_i \approx 6\times10^{-8}\,\mathrm{cm}$. Since $d_i \ll \lambda_{\mathrm{ph}}$, a photon typically traverses $\sim \frac{\lambda_{\mathrm{ph}}}{d_i} \approx 10^2$ ion inertial lengths between scatterings, so the radiation field cannot resolve the current sheet on the scale relevant for reconnection. The large optical depth can still affect particle energization through radiative losses, this effect is already included explicitly in the hadronic efficiency, Eq.~\ref{eq:S_proton}. In Fig.~\ref{fig:cooling_times} we can notice that Inverse-Compton and Bethe-Heitler losses do not dominate over hadronic interactions in the relevant energy range. We therefore conclude that while radiation may affect large-scale flows through viscosity or drag, it does not prevent the formation of fast, plasmoid-mediated reconnection at kinetic scales, where particle acceleration takes place.

We therefore adopt the following physical picture. Proton acceleration takes place within the dense core and is constrained by the Hillas criterion, that will be estimated in the next subsection, which sets the maximum achievable energy based on the size and magnetic field of the system. The proton acceleration that gives rise to pion production is mediated by the magnetic reconnection process itself, a mechanism also considered in \cite{Xiao_2016}. The reconnection regions (current sheets) and outflow jets are characterized by an intrinsically reduced field strength, of the order of $B_r \sim R B_{\rm sat} \sim 10^7$G, where $R$ is the reconnection rate (the reader can find the details in Appendix~\ref{B}, subsection d). Pions, once created within or near these current sheets, are directly injected into the reconnection outflows. They propagate then radially outward until they decay into neutrinos in the outer regions of the merged remnant, where the density drops sharply below the degenerate core value\footnote{The pions that reach the less dense external layers ($\rho \lesssim 100\,\mathrm{g\,cm^{-3}}$) have a better chance of producing detectable high-energy neutrinos; therefore, an estimate of the mean free path of these particles is required. Our post-collision scenario is more catastrophic than the thermodynamic equilibrium shown in Fig.~\ref{fig:profile40}, as turbulence from the collision ejects mass from the envelope, expanding the external layers and further reducing their density profile. In this work, we assume that the high-energy neutrino signal is not significantly attenuated by the density of these external layers. A dedicated simulation of the post-merger hydrodynamics is required to fully confirm this approximation.}.

\subsection{Detectability}

Among the resulting neutrino flavors, muon neutrinos (\(\nu_\mu\) and \(\bar{\nu}_\mu\)) benefit from a significantly larger effective area in the IceCube detector compared to electron neutrinos. Therefore, we concentrate on the muon neutrino signal, avoiding the additional complexity of the multi-step decay chains associated with electron neutrinos.

The available magnetic energy in this region is $\sim 10^{46}$~erg; we must determine the fraction of this energy converted into cosmic rays (CRs) and the hadronic efficiency of neutrino production. In the central region, the magnetic field reaches its maximum value of $B \approx 1.7 \times 10^{10}$~G. Using the Hillas criterion for the maximum proton energy:
\begin{equation}
    E_{p,\max} \approx ZeBR_{\rm m}\beta,
\end{equation}
where $Z = 1$ for protons, $R_{\rm m} \approx 0.01\,R_\odot$, and $\beta = v_A/c$. With an Alfvén velocity $v_A = B/\sqrt{4\pi\rho} \approx 28$~km~s$^{-1}$, we find $E_{p,\max} \approx 3.84 \times 10^{17}$~eV. In hadronic $pp$ interactions, a fraction $\kappa_{pp} \sim 0.5$ of the proton energy is transferred to pions, and each neutrino typically carries $\approx 1/4$ of the parent pion's energy ($E_{\nu} \approx 0.05 E_p$), yielding $E_{\nu,\max} \approx 1.9 \times 10^{16}$~eV.

As explained above, we assumed that hadron acceleration occured inside the remnant, where the magnetic field reaches its maximun value and where Hillas criterion was used. The hadronic efficiency in this region is defined as \citep{Xiao_2016}:
\begin{equation}\label{eq:S_proton}
    \zeta_{\rm CR}(E_{\nu}) = \frac{t^{-1}_{pp} + t^{-1}_{\gamma p} + t^{-1}_{p \alpha}}{t^{-1}_{pp} + t^{-1}_{\gamma p} + t^{-1}_{p\alpha}+ t^{-1}_{\rm ad} + t^{-1}_{\rm syn} + t^{-1}_{\rm IC} + t^{-1}_{BH}},
\end{equation}
where the adiabatic cooling timescale $t_{\rm ad}$ is comparable to the dynamical timescale. Charged pions also experience cooling before decay, leading to an additional suppression of the high-energy neutrino flux. The meson-level suppression factor is defined as:
\begin{equation}
\zeta_{\rm mes}(E_\nu) = \frac{t_{\pi,\rm dec}^{-1}}{t_{\pi,\rm dec}^{-1} + t_{\pi, \rm syn}^{-1} + t_{\pi,\rm had}^{-1}}.
\label{eq:S_mes}
\end{equation}

\begin{figure}
    \centering
    \includegraphics[width=0.95\linewidth]{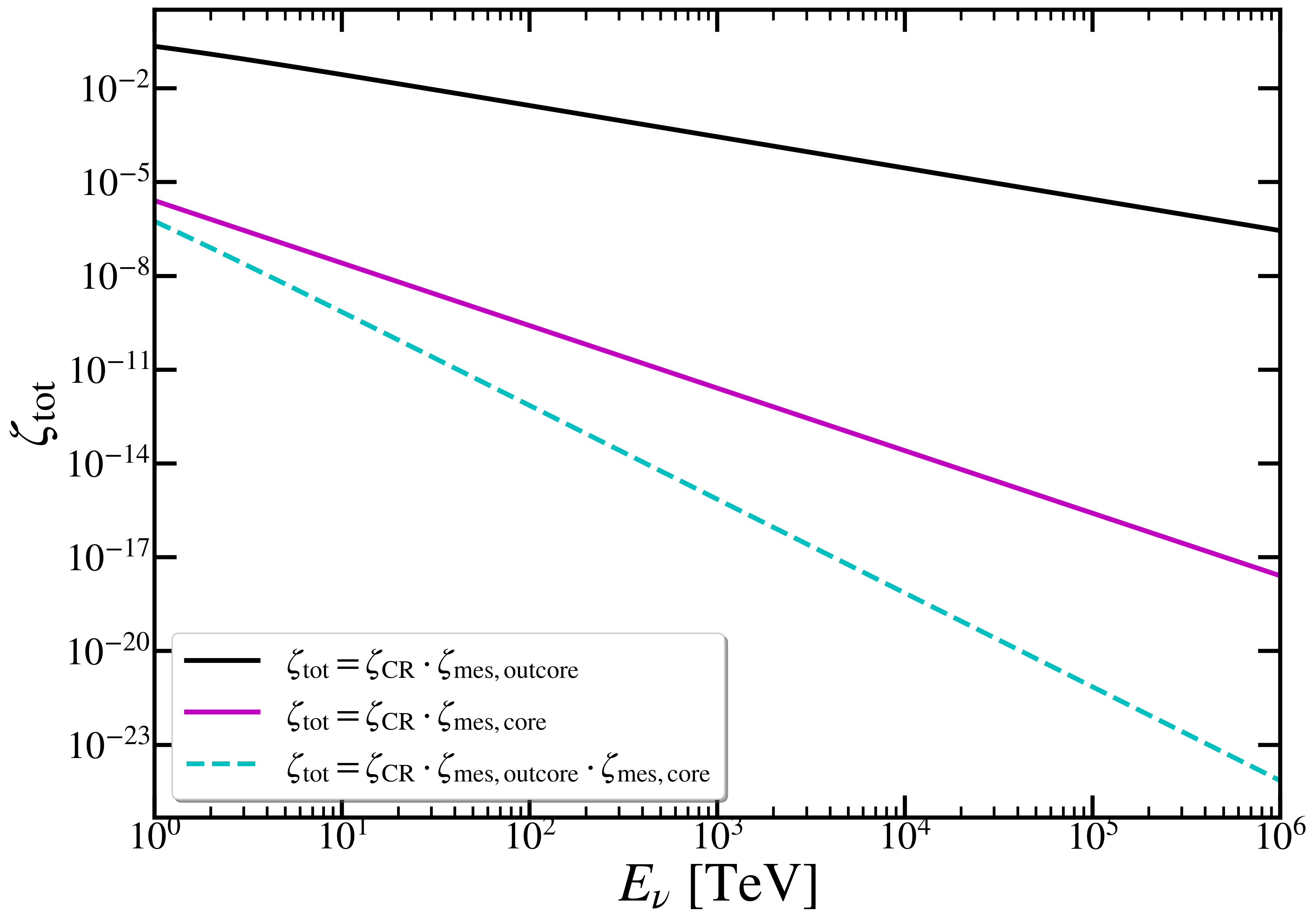}
    \caption{The suppression factor $\zeta_{\rm tot}$ is evaluated for the three terms of Eq.(\ref{eq:zetatotal}). The suppression factor for pions is very sensible to the magnetic field, this is why the process of neutrino emission is more efficient outside the core of the remnant. We can also notice that for higher energies the suppression factor becomes smaller in every cases.}
    \label{fig:zetatotal}
\end{figure}

In the previous section, we conclude that it was not efficient for the pions to interact inside the core of the remnant. However, we are still going to evaluate the suppression factor $\zeta_{\rm mes}$ for three cases, when the pions interact inside the remnant ($\zeta_{\rm mes,core}$), when they interact in the less dense region with the reconnection field and when we have a combination of both, so the reader can compare itself the validity of the model. The overall suppression factor is plotted in Fig.~\ref{fig:zetatotal} and it is determined by: 

\begin{align}\label{eq:zetatotal}
\zeta_{\rm tot}(E_\nu) 
&= \zeta_{\rm CR}(E_\nu)\,
\Bigl( 
  \zeta_{\rm mes,core}(E_\nu) 
  + \zeta_{\rm mes,outcore}(E_\nu) \\ \nonumber
  &+ \zeta_{\rm mes,core}(E_\nu)\,\zeta_{\rm mes,outcore}(E_\nu) 
\Bigr),
\end{align}

As we can notice, the efficiency of pion decays into a detectable neutrino signal comes mainly from the contribution of the second term. From now on, we will refer to this term as $\zeta_{\rm mes}$. The cooling timescales are detailed in Appendix \ref{B} and illustrated in Fig.~\ref{fig:cooling_times}. 

\begin{figure}
    \centering
    \includegraphics[width=0.95\linewidth]{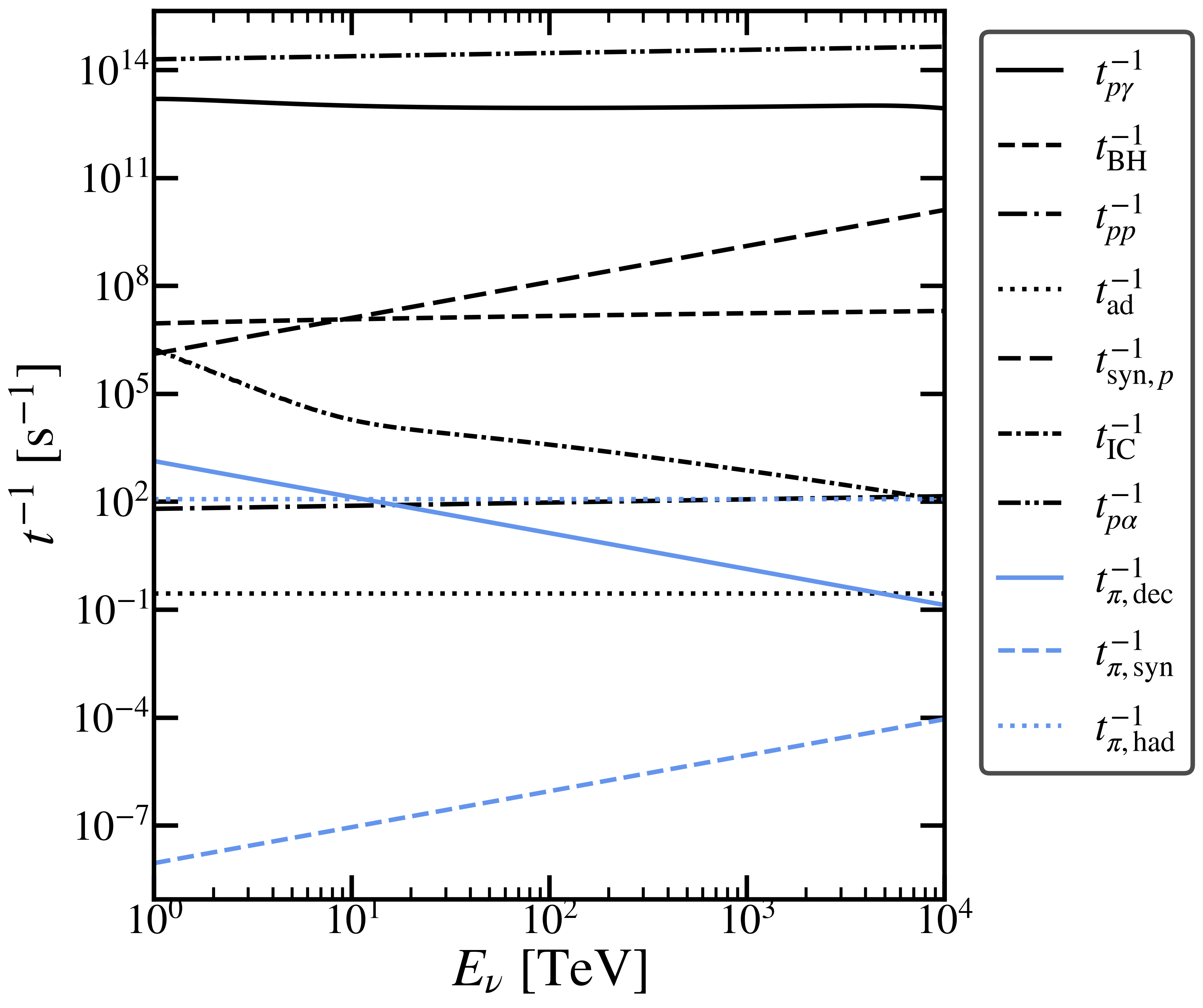}
    \caption{Inverse cooling timescales considered in this work. The black functions represent those asociated to proton interactions inside the core of the remnant. The photohadronic process is the most efficient among the hadronic process that contribute to the high energy neutrino flux ($pp, p\alpha, p\gamma$). On the other hand, the blue plots represent the process associated to pion interactions outside the core of the remnant (we only show this region as it is the dominant term in Eq.(\ref{eq:zetatotal})). We can infer from this graphic that the pion-related suppression factor significantly affects the high-energy neutrino flux, more than the hadronic part.}
    \label{fig:cooling_times}
\end{figure}


\subsection{Source Identification}

Our analysis follows the framework of \cite{Xiao_2016}, though our scenario involves a constant magnetic field and cooling rates that differ by several orders of magnitude. We estimate the maximum distance at which IceCube could detect the signal above 100~TeV using:
\begin{equation}
N_{>100\,\mathrm{TeV}} = \int d\Omega\int_{100\,\mathrm{TeV}}^{E_{\nu,\max}} A_{\rm eff}(E_\nu) \mathcal{F}(E_\nu) \, dE_\nu,
\label{eq:events}
\end{equation}
where the effective area $A_{\rm eff}(E_\nu)$ is taken from \cite{IceCube:2016tpw} and the neutrino flux is:
\begin{equation}\label{eq:flujomuones}
E^2_\nu \mathcal{F}(E_\nu) = \frac{K(1+K)^{-1}\eta_{\rm CR} E_{\rm B} \zeta_{\rm tot}(E_\nu)}{16\pi D^2 \ln \left( \frac{E_{p,\max}}{E_{p,\min}} \right)},
\end{equation}
with $K(1+K)^{-1} \approx 0.14$ \citep{Murase_2016}. For a cosmic ray efficiency $\eta_{CR} = 0.1$, we evaluate the neutrino detection horizon---defined as the distance at which IceCube is expected to detect, on average, one neutrino event ($\mathcal{N} = 1$). We find this horizon to be $\approx 55$~kpc. Muon tracks induced by astrophysical neutrino interactions constitute the main signal category in the search for point-like neutrino sources. Equations (\ref{eq:events}) and (\ref{eq:flujomuones}) account only for this type of neutrino.

To further characterize the observational prospects, Table \ref{tab:neutrino_multiplicity} reports the expected number of detected muon-track neutrino events as a function of the source distance.

\begin{table}[htbp]
\centering
\caption{Expected Neutrino Multiplicity vs. Distance}
\begin{tabular}{cc}
\hline
\hline 
\textbf{Expected Events} & \qquad \textbf{Source Distance} \\
\textbf{(N)} & \textbf{(kpc)} \\
\hline
1 & 55.0 \\
2 & 38.9 \\
3 & 31.7 \\
4 & 27.5 \\
\hline
\end{tabular}
\label{tab:neutrino_multiplicity}
\end{table}

It is crucial to distinguish the detection horizon ($N=1$) from the criteria for unambiguous source identification. For continuous steady-source searches, the atmospheric neutrino background (from pion, kaon, and charmed meson decays) is a formidable obstacle. IceCube \citep{Aartsen_2016} finds only $\sim 500$ cosmic events out of $\sim 350,000$, underscoring its dominance. Thus, neutrino multiplets are required for point-source discovery, as a single muon track cannot be statistically distinguished from the background.

Point-source identification is further challenged by accidental background clustering. Background pair expectations are non-zero (e.g., $\sim 0.8$ for a $0.5^\circ$ cone \citep{MuraseWaxman2016}), implying a false doublet probability of that order. However, false triplets or higher-order multiplets are negligibly small in high-energy track data.

However, erythrohenosis events constitute a transient multimessenger source rather than a steady neutrino emitter. The high-energy neutrino emission is strictly confined to the interval between core contact and helium ignition, corresponding to a duration of approximately 196~s in our model. In this context, the observational strategy could be enhanced by a temporally and spatially constrained transient search.

The preceding inspiral of the degenerate cores is expected to produce a distinct gravitational-wave (GW) signal in the LISA band, which acts as a precise temporal trigger ($\sim$1 year before cores contact) and provides a preliminary localization area. A subsequent electromagnetic (EM) transient similar to a Luminous Red Novae (LRNe) can further pinpoint the host galaxy. A more detailed description of this multimessenger phenomena can be found in \citep{seoane2026erythrohenosis}.

\subsection{Diffuse Neutrino Flux}
\begin{figure}
    \centering
    \includegraphics[width=0.95\linewidth]{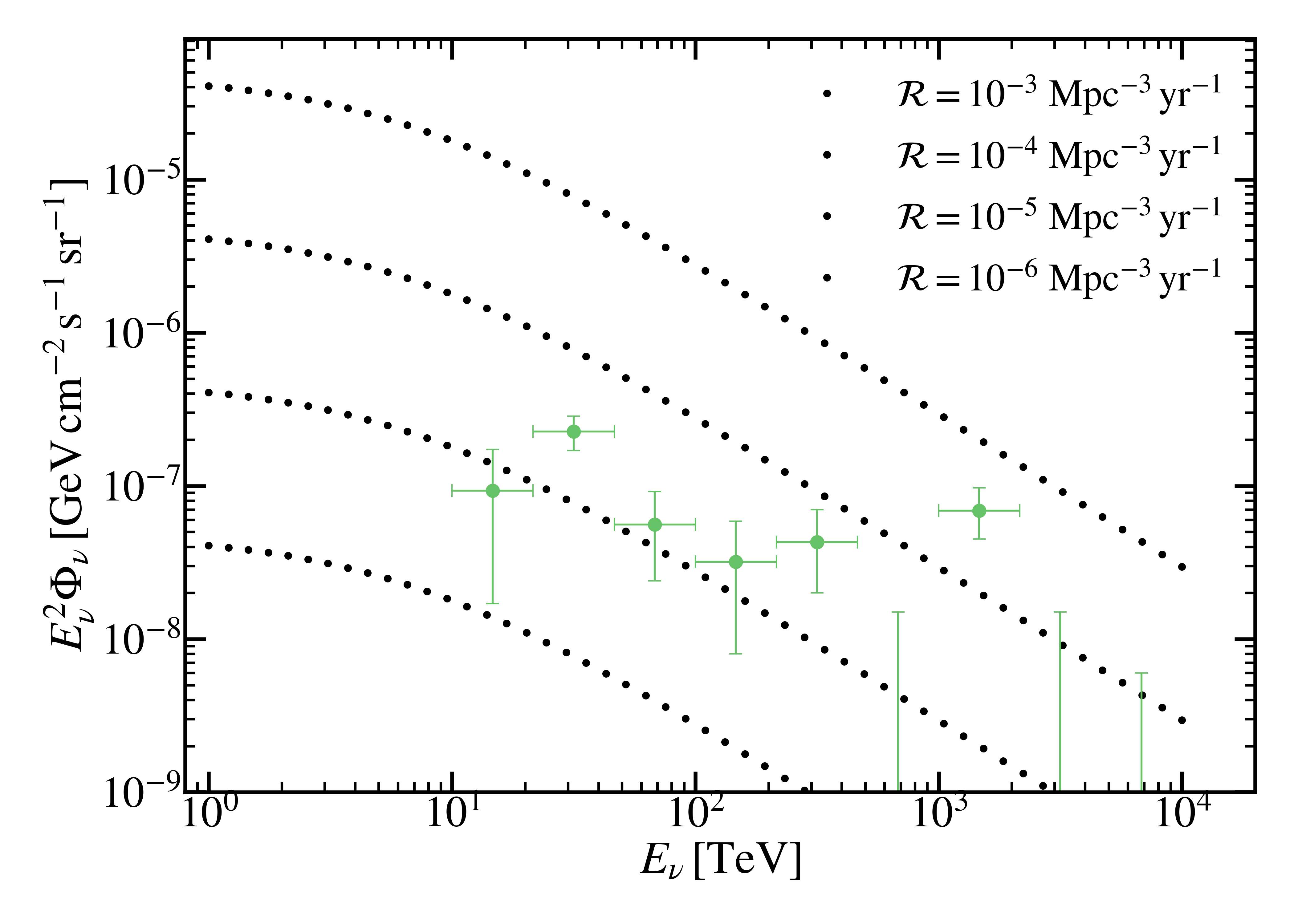}
    \caption{Diffuse neutrino flux (all-flavor) predicted by our model for various collision rates $\mathcal{R}$. The green data points represent IceCube observations \citep{Aartsen_2015}. We can notice that the signal at higher energies decreases, and this makes sense because of the suppression factors estimated in Fig.\ref{fig:zetatotal}.}
    \label{fig:icecube}
\end{figure}

The diffuse neutrino flux (per flavor) is given by:
\begin{equation}
E_{\nu}^{2} \Phi_{\nu_{i}} = \frac{c}{4\pi H_{0}} 
\int_{0}^{z_{\mathrm{max}}} \frac{\epsilon_{\nu} Q_{\epsilon_{\nu_{i}}} \, S(z)}
{(1+z)^{2} \sqrt{\Omega_{M}(1+z)^{3} + \Omega_{\Lambda}}} \, dz, \label{eq:flux}
\end{equation}
where
\begin{equation}
\epsilon_{\nu} Q_{\nu_{\mathrm{tr}}} = \frac{K}{4(1+K)} \,
\frac{\mathcal{R} E_{\rm CR}}{\ln(\epsilon_{p,\max} / \epsilon_{p,\min})} \, \zeta_{\rm tot}(\epsilon_{\nu}) \label{eq:comoving}
\end{equation}
represents the differential neutrino emission rate (energy per unit time per unit comoving volume) for a given flavor. Here, $\mathcal{R}$ denotes the rate of red giant core collisions. While several studies have attempted to constrain this rate (see \cite{seoane2026erythrohenosis} for a summary), it remains a highly uncertain parameter. The galaxies density varies along all the universe, and also the amount of globular clusters of each one. In Fig.(1) of \citep{Burkert2010} it is shown that the number of globular clusters can increase by several orders of magnitude. Consequently, we adopt some possible values in Fig.~\ref{fig:icecube}.

In Eq.~(\ref{eq:flux}), $Q_\nu$ is defined in the comoving volume (the notation $\epsilon_\nu$ refers to the energy in the comoving frame), $H_0$ is the Hubble constant, and $\Omega_M$ and $\Omega_\Lambda$ are the present-day density parameters for matter and dark energy, respectively. For a detailed discussion of these parameters and the source evolution with redshift $S(z)$, we refer the reader to \cite{Xiao_2016}.

The high-energy neutrino flux measured by IceCube between $\sim 10$~TeV and a few PeV provides a unique probe of extragalactic neutrino source populations. Despite over a decade of data, the origin of this flux remains a subject of debate. Our work demonstrates that the merger of degenerate red giant cores can successfully reproduce the observed IceCube diffuse flux under physically motivated assumptions, adopting event rates in globular clusters that are consistent with the plausible ranges previously established in the literature. Notably, similar to white dwarf mergers, these sources are optically thick to high-energy gamma-rays, characterizing them as "hidden" accelerators—a result consistent with the lack of associated gamma-ray detections by \textit{Fermi}-LAT.

\section{Helium detonation likelihood}\label{sec:6}
As previously noted, the calculated temperature rise is sufficient to lift the electron degeneracy, likely triggering a helium flash. Since the core density and pressure remain high, such an event is highly probable. It is important to refine the terminology regarding the nature of this ignition. While the title and preceding sections use the term "detonation" for simplicity, the physical conditions warrant a more nuanced discussion. Because the collision occurs at subsonic velocities, coalescence proceeds via quasi-static compression rather than a strong shock. Therefore, the initial burning front is expected to propagate as a \textit{deflagration}—a subsonic flame front mediated by thermal conduction and turbulent transport.

However, the extreme environment allows for a more complex evolution. The turbulent dynamo that amplifies the magnetic field also generates vigorous small-scale turbulence, which can wrinkle the burning front and increase its surface area, potentially accelerating the deflagration. In such magnetized, turbulent media, a deflagration-to-detonation transition (DDT) may occur, leading to a runaway supersonic detonation. Such transitions are well-studied in thermonuclear supernovae, where magnetic fields and turbulence are critical triggers. In the erythrohenosis scenario, the strongly magnetized environment and rapid heating could facilitate a DDT, particularly in the outer, less-degenerate layers or magnetized outflows.

While we retain ``detonation'' in the title to reflect the explosive outcome, we emphasize that ignition likely begins as a deflagration. The possibility of a DDT remains an intriguing question for future multidimensional simulations. This distinction does not alter our primary conclusions, as the total energy release and nuclear activation are governed by the global thermodynamic evolution rather than the specific details of flame propagation.

Direct observations of a standard helium flash are precluded because the resulting luminosity ($\sim 10^{11}~L_\odot$) is consumed in lifting the core degeneracy. However, in this collisional scenario, the triple-alpha process triggers the production of $^{18}\mathrm{F}$ from $^{14}\mathrm{N}$. The neutrinos from $^{18}\mathrm{F}$ decay could provide a direct observational signature of the flash, allowing us to probe this phenomenon for the first time.

Furthermore, observing such events could validate models of thermal neutrino cooling. Although theoretical models for processes like plasmon decay are well-established, they have never been directly observed in stars. As shown by \cite{Capelo_2023}, thermal neutrinos determine whether the helium flash occurs as a single event or a series of subflashes, each with distinct asteroseismic signatures. With the upcoming PLATO mission expected to launch in late 2026 \citep{rauer2024platomission}, high-precision asteroseismology may soon provide indirect tests for these models. The erythrohenosis scenario offers a complementary high-energy window into these fundamental stellar processes.

\subsection{Magnetic Field Effects on the \(^{14}\mathrm{N}(\alpha,\gamma)^{18}\mathrm{F}\) Reaction Rate}

The detection of these neutrinos has broader implications than previously discussed. Since the neutrino flux from this reaction depends on the abundance of $^{14}\mathrm{N}$ in the stellar core, it could serve as a probe of elemental composition, helping to refine the chemical profile of different stellar regions. Additionally, $^{18}\mathrm{F}$ neutrinos are the only species whose production is expected to scale linearly with metallicity regardless of stellar mass \citep{Capelo_2023}. When paired with independent mass determinations, this property could be invaluable for discerning the metallicity of similar objects and establishing metallicity maps of observed sources.

As discussed, the rapid heating of the degenerate helium core to temperatures exceeding \(10^{8}\,\mathrm{K}\) triggers not only the triple-alpha process but also the deflagration of the \(^{14}\mathrm{N}\) accumulated during prior CNO cycling. This nitrogen flash proceeds via the reaction:
\begin{equation}
^{14}\mathrm{N} + \alpha \rightarrow ^{18}\mathrm{F} + \gamma,
\end{equation}
followed by the $\beta^{+}$ decay of the radioactive \(^{18}\mathrm{F}\) nucleus:
\begin{equation}
^{18}\mathrm{F} \rightarrow ^{18}\mathrm{O} + e^{+} + \nu_e.
\end{equation}
This decay emits an electron neutrino with a maximum energy of \(0.633\,\mathrm{MeV}\) and an average energy of \(\langle E_\nu \rangle \approx 0.382\,\mathrm{MeV}\).

\begin{figure}
    \centering
    \includegraphics[width=0.95\linewidth]{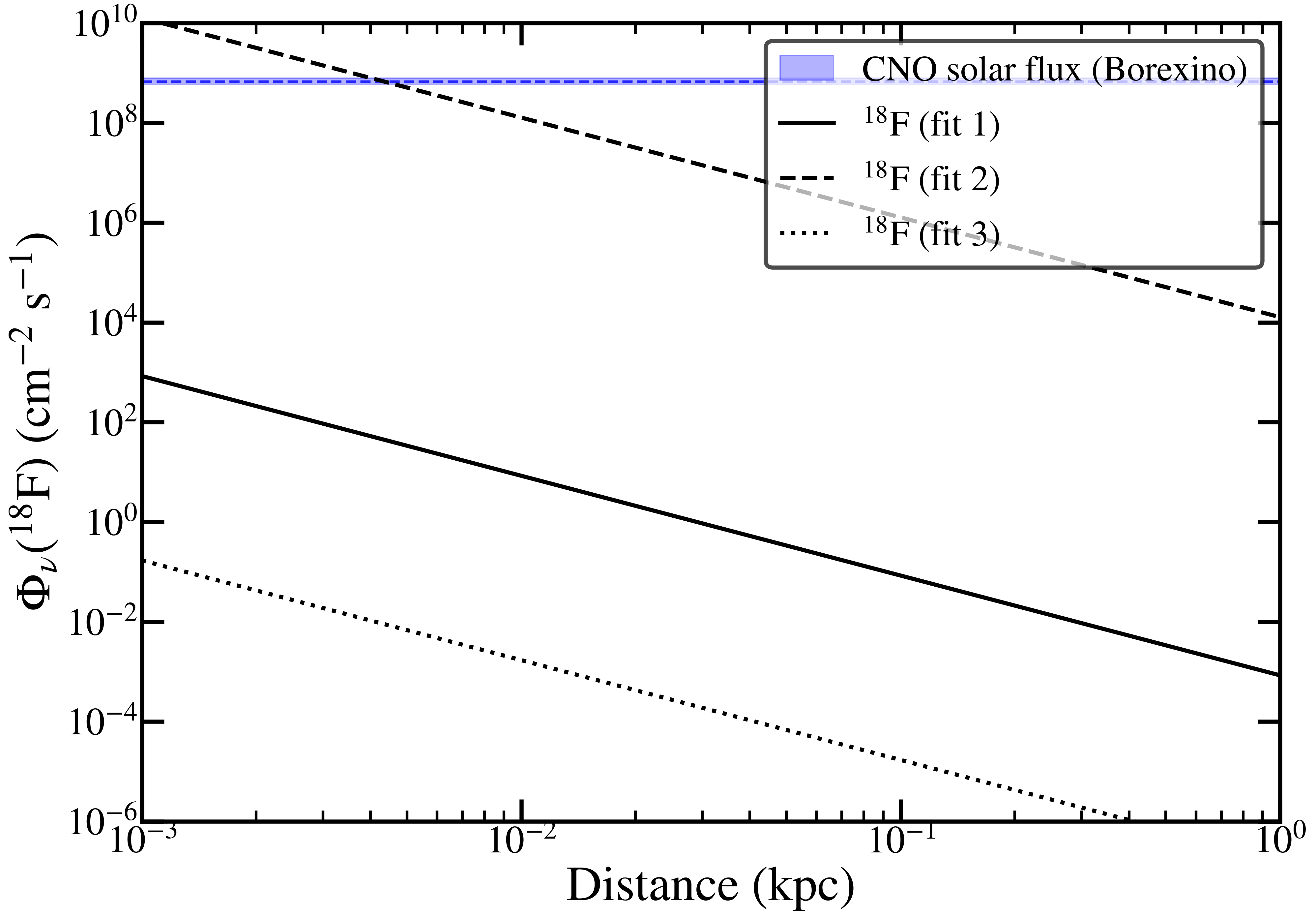}
    \caption{Dependence of the neutrino flux on distance up to 1~kpc (for reference, the distance to the Galactic Center is $\approx 8$~kpc) using various fitting parameters from \cite{Iliadis_2010} (black curves), assuming $f \approx 1$. The flux is compared with the total solar CNO neutrino flux (blue band, $\Phi_{\rm CNO} = 6.7^{+1.2}_{-0.8} \times 10^8$ cm$^{-2}$ s$^{-1}$) \citep{Basilico_2023}, which occupies the same energy window. The signal from the fit 2 would be equal to the CNO neutrino flux at 4.4 pc. This signal would represent a minor perturbation except in the most favorable cases where the source is very close.}
    \label{fig:f18}
\end{figure}

The reaction rate is defined as:
\begin{equation}\label{eq:reactionrate}
    r = n_{\rm ^{14}N} n_{\rm ^{4}He} f \langle \sigma v \rangle, 
\end{equation}
where $n_i$ is the number density for each isotope, $f$ is the screening factor, and $\langle \sigma v \rangle$ is the thermally averaged cross section. Using the JINA Reaclib library and the work of \cite{Iliadis_2010}, we obtain the neutrino flux shown in Fig.~\ref{fig:f18}.
It should be noted that the relative abundance of $^{18}{\rm F}$ is not explicitly tracked in our calculations, since the SPH simulations does not take into account the physics after the two cores collide. Fig.~\ref{fig:f18} shows the rate of the $^{14}{\rm N}(\alpha,\gamma)^{18}{\rm F}$ reaction rather than the subsequent $\beta^+$ decay of $^{18}{\rm F}$. This approximation is justified because the half-life of $^{18}\mathrm{F}$ ($t_{1/2}=109.7~\mathrm{min}\simeq6.6\times10^{3}~\mathrm{s}$) is shorter than the characteristic nitrogen-burning timescale ($\tau_{14\alpha}\simeq5\times10^{4}~\mathrm{s}$), the isotope rapidly approaches a quasi-steady-state abundance. In this regime, the neutrino production rate is approximately equal to the $^{14}\mathrm{N}(\alpha,\gamma)^{18}\mathrm{F}$ production rate, validating the use of the latter as a proxy for the observable neutrino flux, under the physical conditions considered here. Consequently, the neutrino production rate associated with the decay of $^{18}{\rm F}$ is effectively determined by the slower
$\alpha$-capture reaction.

The resulting neutrino flux is comparable to the solar CNO flux only if the source is exceptionally close to Earth (we refer the reader to Fig.1 of \citep{CNOwindow} to compare the energy window for these neutrino signal.). In the work of \citep{Irene}, they point out that there are no stellar candidates approaching the tip of the red giant branch within 10 pc, and in \citep{Serenelli_2005} they estimate the $^{18}\rm F$ neutrino flux at this distance as only a 10\% of the CNO neutrino flux. However, these works search for individual stars that naturally evolve up to helium flash. We are dealing with collisions of red giant stars that accelerate their detonation due to the inestabilities caused by the shock betweeen degenerate cores. This kind of stars can be found in globular clusters whitin 10 pc. Of course, we have to take into account that the collision rate is extremelly low, but yet, it is worthy to analyse it.

As the neutrino flux scales as \(1/d^2\) and taking the CNO-equivalent reference flux at \(4.4\,\mathrm{pc}\), the expected relative flux for nearby globular clusters is
\[
\frac{\Phi(d)}{\Phi_{\mathrm{CNO}}(4.4\,\mathrm{pc})}=\left(\frac{4.4}{d}\right)^2.
\]
Under this scaling, M4 at \(2.21\,\mathrm{pc}\) would yield about 4 times the reference flux, NGC 6397 at \(2.39\,\mathrm{pc}\) about 3.39, M22 at \(3.19\,\mathrm{pc}\) about 1.90, and 47 Tucanae at \(3.99\,\mathrm{pc}\) about 1.21, all of them at or above the reference level. By contrast, more distant clusters such as M5 at \(7.52\,\mathrm{pc}\), M15 at \(10.31\,\mathrm{pc}\), M3 at \(10.40\,\mathrm{pc}\), and M2 at \(11.50\,\mathrm{pc}\) would contribute only about \(34\%\), \(18\%\), \(18\%\), and \(15\%\) of the reference flux, respectively. This indicates that only the nearest globular clusters would approach or exceed a CNO-like flux level defined at \(4.4\,\mathrm{pc}\).

Currently, measuring neutrinos from $^{18}\mathrm{F}$ decay remains challenging due to the significant background from solar neutrinos. Nevertheless, future low-background detectors may eventually be able to observe this signal. In this context, the Jinping Neutrino Experiment has been proposed as a particularly promising low-energy neutrino observatory. Its design foresees a fiducial mass of approximately 2~kt for neutrino--electron elastic scattering studies, while benefiting from an exceptionally low cosmic-ray muon flux of $(2.0 \pm 0.4)\times10^{-10}\,\mathrm{cm^{-2}\,s^{-1}}$, about two hundred times lower than that measured at Borexino. In addition, the great depth of the China Jinping Underground Laboratory and its large distance from nuclear power plants substantially reduce cosmogenic and reactor-neutrino backgrounds, making it an attractive facility for the detection of rare low-energy neutrino signals \citep{Beacom2017Jinping}.

The saturated magnetic field \(B_{\mathrm{sat}} \approx 1.77 \times 10^{10}\,\mathrm{G}\) present at the onset of the helium flash motivates an investigation into its impact on the \(^{14}\mathrm{N}(\alpha,\gamma)^{18}\mathrm{F}\) rate. Following the formalism of \cite{Famiano_2020} for nuclear screening in magnetized plasmas, we find that magnetic enhancement is negligible in this scenario. In the post-merger core (\(\rho \approx 10^{6}\,\mathrm{g\,cm^{-3}}\), \(T \approx 2 \times 10^{8}\,\mathrm{K}\)), the plasma is in the Thomas-Fermi (TF) weak screening regime. The relevant dimensionless parameters:
\begin{equation}
\frac{\hbar \omega_c}{k_B T} \approx 0.0118, \quad \frac{\hbar \omega_c}{E_{\mathrm{Fermi}}} \approx 0.00125,
\end{equation}
indicate that Landau quantization of the electron density is minimal. The leading magnetic correction to the TF screening length is on the order of \(5.2 \times 10^{-7}\). The zero-field TF length is $\lambda_{\mathrm{TF}} \approx 4.5 \times 10^{-10}$~cm, yielding a static screening factor $f_0 \approx 1.30$. Magnetic corrections increase this by only $\sim 3 \times 10^{-5}\%$. Furthermore, dynamic screening effects are negligible because the number of occupied Landau levels is large (\(N_{\mathrm{occ}} \sim 800\)), causing the plasma response to revert to the unmagnetized case. Thus, $^{18}\mathrm{F}$ production proceeds at the standard rate.

\subsection{CNO activation post-detonation}

The timescale for the CNO cycle is large compared to the high-energy neutrino production timescales estimated for the pre-flash phase. However, immediately following the flash, the core degeneracy is lifted, and the remnant transitions into a hot, non-degenerate helium-burning zone with a convective core. While the resulting neutrinos from the $^{14}\rm N(\alpha,\gamma)^{18}\rm F$ are search as the most important window for proving the helium flash in normal red giant stars, in our particular case we have a non-zero abundance of hydrogen in the degenerate core of the remnant. Mixed protons then serve as fuel for the CNO cycle, also producing neutrinos through:
\begin{align*}
&^{13}\mathrm{N} \rightarrow {}^{13}\mathrm{C} + e^{+} + \nu_{e} \quad (E_{\nu,\max} = 1.199\,\mathrm{MeV}, \text{ CNO-I}), \\
&^{15}\mathrm{O} \rightarrow {}^{15}\mathrm{N} + e^{+} + \nu_{e} \quad (E_{\nu,\max} = 1.732\,\mathrm{MeV}, \text{ CNO-I}), \\
&^{17}\mathrm{F} \rightarrow {}^{17}\mathrm{O} + e^{+} + \nu_{e} \quad (E_{\nu,\max} = 1.740\,\mathrm{MeV}, \text{ CNO-II}), \\
&^{18}\mathrm{F} \rightarrow {}^{18}\mathrm{O} + e^{+} + \nu_{e} \quad (E_{\nu,\max} = 0.633\,\mathrm{MeV}, \text{ CNO-III}).
\end{align*}

\begin{figure}
    \centering
    \includegraphics[width=0.95\linewidth]{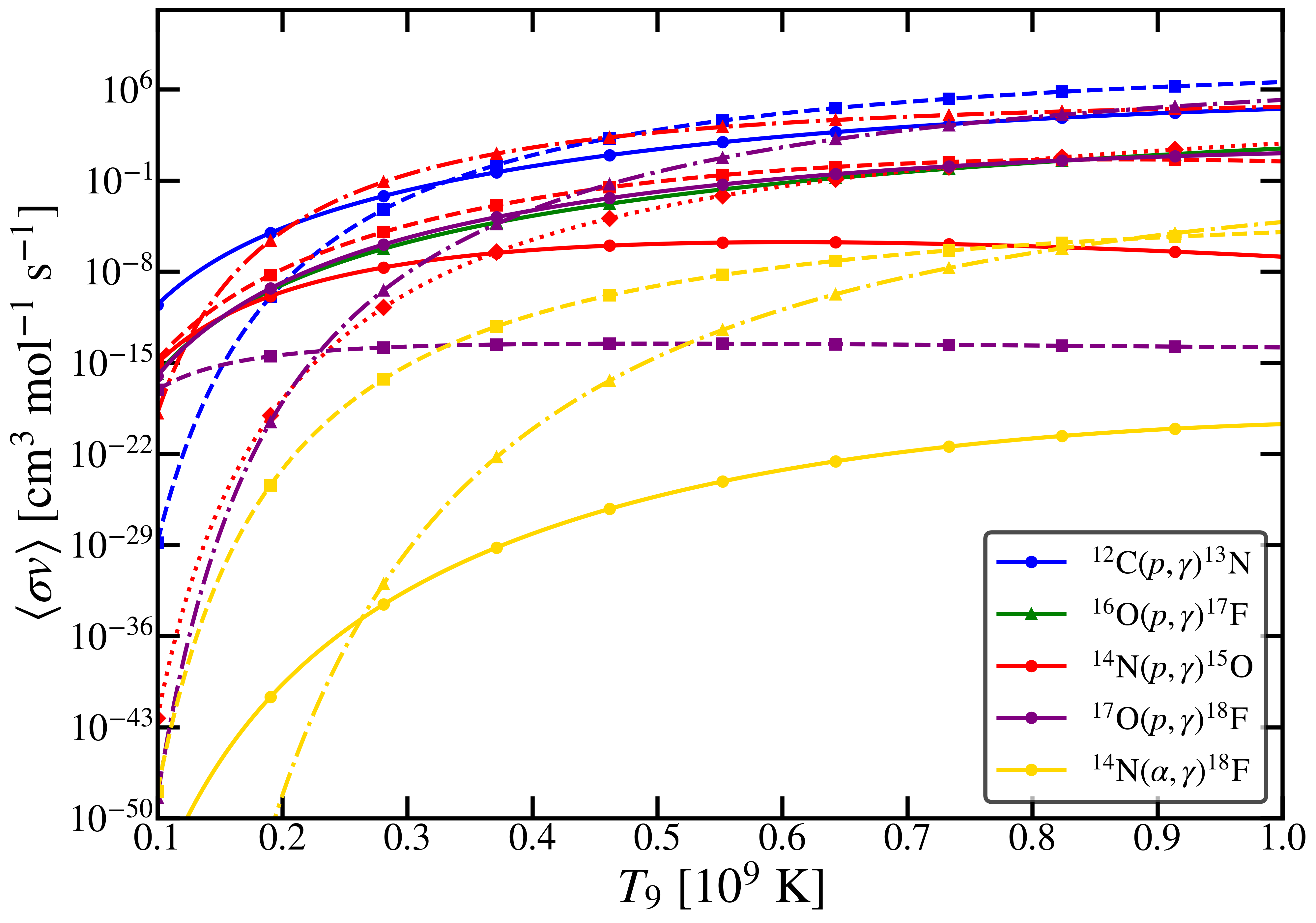}
    \caption{The blue fits correspond to the $^{12}\text{C} (p,\gamma)^{13}\text{N}$ reaction rate \citep{Li2010}, the green fit to the $^{16}\text{O} (p,\gamma) ^{17}\text{F}$ reaction rate \citep{Iliadis2008}, the red fits to the $^{14}\text{N}(p,\gamma)^{15}\text{O}$ reaction rate \citep{Imbriani2005} and the purple fits to the $^{17}\text{O}(p,\gamma)^{18}\text{F}$ reaction rate \citep{Iliadis2010}. All of these data can be found in JINA Reaclib library. We can also notice that the \(^{14}\mathrm{N}(\alpha,\gamma)^{18}\mathrm{F}\) rate (yellow fits) is the slowest one.}
    \label{fig:CNOsigma}
\end{figure}

A detailed analysis of this decays and its consequent neutrino flux is a complex work and should be address in a future work. First of all, abundances evolution is not a simple task when we are dealing with such turbulent scenarios. Dedicated simulations must be used in order to have good estimations of this signal. However, we would like to make some comparisons. Assuming that the abundances of Fig.~\ref{fig:profile40abundancias} are approximatelly the same, in Fig.~\ref{fig:CNOsigma} it is shown the thermally averaged reaction rate ($<\sigma v>$) for Temperatures from 0.1-1.0 $\times 10^9$ K for the parent reactions of the CNO cycle. We have also plotted the three fits of the reaction $^{14}\mathrm{N}(\alpha,\gamma)^{18}\mathrm{F}$ shown in Fig.~\ref{fig:f18}. It is interesting to notice that $^{14}\rm N$ is the bottleneck in the CNO cycle. When the helium flash is approaching, this reaction and its consequent $^{18}\rm F$ decay gives a maximum neutrino luminosity of 10${^8}$ times the solar bolometric luminosity \citep{Serenelli_2005}. Despite this last reaction is the slowest one, the relative abundance of $^{14}\rm N$ and $^{4}\rm He$ over $^{12}\rm C$, $^{16}\rm O$, $^{17}\rm O$ and, of course $^{1}\rm H$, makes the reaction rate very significative (see Eq.\ref{eq:reactionrate}).

If we take into account our previous results and keep in mind the relative abundances, we can think that the Solar neutrino flux background is still quite a problem for this CNO signal if we are dealing with extragalactic sources. Nevertheless, the activation of an hydrogen core has profound evolutionary consequences. Spectroscopically, this manifests as anomalies in light elements (e.g., C/N and O/Na anticorrelations) frequently observed in globular clusters \citep{1999ApJdeepmixing}. Asteroseismically, the inversion in mean molecular weight from hydrogen ingestion would create characteristic mode frequency shifts detectable by the future PLATO mission. Additionally, the Cameron-Fowler mechanism could lead to lithium enrichment, explaining lithium-rich red clump stars. Finally, helium enhancement from hydrogen fusion alters the star's horizontal branch position, potentially explaining the spread of horizontal branch colors in globular clusters. Erythrohenosis thus provides a unique framework for understanding the physics of extreme mixing and the anomalous abundance patterns in dense stellar populations.

\section{Summary and Discussion}\label{sec:7}

We have investigated the coalescence of two degenerate helium cores from low-mass red giant stars in a kind of dense stellar environment, globular clusters—a process we term \textit{erythrohenosis}—identifying these events as a novel class of transient high-energy neutrino sources. Using stellar models generated with MESA and SPH simulations of the final inspiral phase, we characterized the thermodynamic evolution of the merging cores and developed a model for hydrogen trapping during the collision. Our main findings are as follows.

The merger traps approximately $10^{-14}$ M$_\odot$ of hydrogen between the degenerate cores, yielding a post-mixing proton density of $n_p \approx 10^{17}~\text{cm}^{-3}$ in the remnant. Simultaneously, a turbulent dynamo amplifies the magnetic field to a saturation value of $B_{\rm sat} \approx 10^{10}$~G well before the merger completes. The gravitational energy release, $\Delta U \approx 4.28 \times 10^{49}$~erg, heats the core to $T_f \approx 5.3 \times 10^8$~K. This exceeds the helium ignition threshold ($T_{\rm ign} \sim 10^8$~K), thereby accelerating the star's evolution. Helium ignites at $t_{\rm ign} \approx 196$~s after first contact, well before the completion of the merger ($\tau_{\rm m} \approx 10^3$~s), ensuring that the helium flash occurs within a strongly magnetized environment.

Following the formalism of \citet{Famiano_2020}, we evaluated the impact of the saturated magnetic field on the $^{14}\text{N}(\alpha,\gamma)^{18}\text{F}$ reaction. The static Thomas-Fermi screening factor, $f_0 \approx 1.30$, is enhanced by less than $10^{-4}\%$, and dynamic screening effects are negligible due to the large number of occupied Landau levels. Thus, the production of $^{18}\text{F}$ proceeds at essentially the same rate regardless of the magnetic field strength, and its subsequent $\beta^+$ decay provides a clean, field-insensitive signature of the helium flash.

Protons accelerated via magnetic reconnection produce high-energy neutrinos through $pp$ interactions, $p\alpha$ interactions, and the $p\gamma$ process. Incorporating all relevant cooling processes, we computed the total suppression factor $\zeta_{\rm tot}(E_\nu)$. For a total cosmic-ray energy $E_{\rm CR} \approx 10^{46}$~erg, the expected number of IceCube events above 100~TeV from a single merger at 55~kpc is $\approx 1$. Furthermore, the diffuse neutrino flux from the cosmic population of such events—assuming literature-based rates and source evolution tracing the star formation history—contributes significantly to the IceCube signal. Our results establish red giant core mergers as a viable and promising source class for the IceCube astrophysical neutrino flux.

The $^{18}\text{F}$ signal, unaffected by the magnetic field and varying linearly with metallicity, offers a unique tool for mapping chemical abundances in dense stellar systems \citep{Capelo_2023}. PeV-scale neutrinos probe particle acceleration in magnetized, high-density environments. Unlike the GRB fireball model, where protons interact with a relativistic radiation field, the erythrohenosis environment is characterized by a subsonic regime where pressure forces efficiently redistribute momentum. Consequently, flux suppression due to the synchrotron cooling of pions and muons is less severe than in choked jet or high-Lorentz-factor environments. Furthermore, the coincident arrival of neutrinos and gravitational waves from a nearby event could test the weak equivalence principle and constrain the limiting speed of neutrinos with an accuracy orders of magnitude beyond current solar or supernova constraints.

Beyond neutrino astronomy, erythrohenosis events have broader implications for stellar evolution and nucleosynthesis. CNO cycle activation in the hydrogen-contaminated core produces distinct abundance patterns, such as the C/N and O/Na anticorrelations observed in globular cluster red giants \citep{1999ApJdeepmixing}. The helium produced during hydrogen ingestion is subsequently dredged to the surface, influencing the star's position on the horizontal branch; helium-enhanced stars evolve toward hotter temperatures, populating the "blue tail." The observed spread in horizontal branch colors within globular clusters may thus encode the efficiency of internal hydrogen mixing.

While detecting single events at cosmological distances remains challenging, next-generation detectors like IceCube-Gen2 and KM3NeT will offer the necessary sensitivity. Stacking analyses targeting galactic nuclei and globular clusters could further enhance the signal-to-noise ratio. Moreover, erythrohenosis provides a compelling framework for understanding Luminous Red Novae (LRNe). As concluded in \cite{seoane2026erythrohenosis}, these core mergers are energetic transients whose output is reprocessed by the dynamically ejected envelope, resulting in the characteristic red optical and infrared emission of "supernova impostors."

Finally, the inspiral and merger produce a gravitational wave (GW) signal within the LISA frequency band. Unlike vacuum inspirals, the orbital evolution is dominated by gas drag, imprinting a unique "smoking-gun" signature: a Time-Varying Apparent Chirp Mass (TVACM). The detection of such a signal by LISA would provide an unprecedented probe of common-envelope hydrodynamics. Future research must address the fate of the magnetic field post-flash and the possibility of $r$-process nucleosynthesis in more massive mergers. In conclusion, erythrohenosis events represent a previously underappreciated class of multimessenger sources, offering a window into magnetized helium detonations and the chemical evolution of dense stellar systems.

\section{Acknowledgments}
This work is dedicated to Fernando Guzmán Martínez, who facilitated our collaboration and sadly passed away before its completion. A pillar of his home institute and a respected name worldwide, his suggestions and motivation were instrumental to this project even if he could not see its conclusion.

\appendix
\section{Lagrangian Form of the Induction Equation in Ideal and Incompressible Magnetohydrodynamics} \label{A}

The induction equation is given by
\begin{equation}
\frac{\partial \mathbf{B}}{\partial t} = \nabla \times (\mathbf{v} \times \mathbf{B}) + \lambda \nabla^2 \mathbf{B},
\end{equation}
where ($\mathbf{B}$) is the magnetic field, ($\mathbf{v}$) is the fluid velocity (taken here as the relative velocity of the collision), and ($\lambda$) denotes the magnetic diffusivity. In the ideal-conductor limit, ($\lambda \to 0$), this equation reduces to
\begin{equation}
\frac{\partial \mathbf{B}}{\partial t} = \nabla \times (\mathbf{v} \times \mathbf{B}).
\end{equation}
Using the standard vector identity
\begin{equation}
\nabla \times (\mathbf{v} \times \mathbf{B}) = (\mathbf{B} \cdot \nabla)\mathbf{v} - (\mathbf{v} \cdot \nabla)\mathbf{B} - \mathbf{B}(\nabla \cdot \mathbf{v}) + \mathbf{v}(\nabla \cdot \mathbf{B}),
\end{equation}
and imposing the solenoidal constraint on the magnetic field,
\begin{equation}
\nabla \cdot \mathbf{B} = 0,
\end{equation}
the induction equation may be written as
\begin{equation}
\frac{\partial \mathbf{B}}{\partial t} + (\mathbf{v} \cdot \nabla)\mathbf{B} = (\mathbf{B} \cdot \nabla)\mathbf{v} - \mathbf{B}(\nabla \cdot \mathbf{v}).
\end{equation}

The left-hand side of this expression corresponds to the material (or convective) derivative,
\begin{equation}
\frac{D}{Dt} \equiv \frac{\partial}{\partial t} + \mathbf{v} \cdot \nabla,
\end{equation}
and the induction equation may therefore be written in Lagrangian form as
\begin{equation}
\frac{D\mathbf{B}}{Dt} = (\mathbf{B} \cdot \nabla)\mathbf{v} - \mathbf{B}(\nabla \cdot \mathbf{v}).
\end{equation}

This expression makes explicit that, in the ideal limit, the magnetic field is advected by the flow and evolves according to the plasma motion. In the special case of an incompressible flow ($\nabla \cdot \mathbf{v} = 0$) \citep{FluidMechanics1959}, the evolution of the magnetic field is governed solely by the stretching of magnetic field lines by velocity gradients parallel to $\mathbf{B}$. This is justified by analyzing the mass conservation equation:
\begin{equation}
    \frac{\partial{\rho}}{\partial{t}} + \nabla \cdot(\rho\mathbf{v}) = 0.
\end{equation}
When the fluid is incompressible, $\rho \approx \text{const}$, which is the case here as demonstrated by Eq.~(\ref{eq:lowdensityvariation}). Consequently, we conclude that $\nabla\cdot\mathbf{v} = 0$.

\section{Proton and Pion Cooling Times}\label{B}

In both the $pp$ and $p\gamma$ channels, the neutrinos observed at Earth carry a fixed fraction of the parent proton energy. For charged-pion decay chains $\pi^+ \to \mu^+ \nu_\mu \to e^+ \nu_e \bar{\nu}_\mu \nu_\mu$, each neutrino typically receives a fraction $\sim 5\%$ of the initial proton energy, such that:
\begin{equation}
  E_\nu \simeq 0.05\,E_p \quad \Rightarrow \quad E_p \simeq 20\,E_\nu.
\end{equation}
Consequently, any characteristic proton energy scale $E_{p,{\rm br}}$ (for example, where $t_{\rm syn}(E_p) = t_{\rm dyn}$ or $t_{p\gamma}(E_p)=t_{\rm acc}(E_p)$) translates into a corresponding neutrino spectral break at:
\begin{equation}
  E_{\nu,{\rm br}} \simeq 0.05\,E_{p,{\rm br}}.
\end{equation}
In the following discussion regarding the positions of spectral features in the neutrino flux, we systematically apply the mapping $E_p \simeq 20 E_\nu$ to express the relevant timescales as a function of the observable neutrino energy. To compute the neutrino spectrum at energies $E_\nu \gtrsim 1~\mathrm{TeV}$, we define the relevant cooling timescales and evaluate them in the dense magnetized helium core of the remnant for hadronic interactions, and in the less dense region of the remnant for pion decays, as this combination was the more efficient one for the neutrino signal. The suppression factors for the neutrino luminosity are then expressed as explicit functions of the proton energy $E_p$ and subsequently related to the neutrino energy $E_\nu$.

\subsubsection{Photomeson cooling time}

The photomeson cooling time $t_{p\gamma}(E_p)$ is defined in the plasma rest frame as the inverse of the interaction rate of a proton with Lorentz factor $\gamma_p = E_p/(m_p c^2)$ moving through an isotropic photon field with spectral number density $n_\gamma(\varepsilon)$:
\begin{equation}
t_{p\gamma}^{-1}(E_p) = c \int d\varepsilon\, n_\gamma(\varepsilon) \left\langle \sigma_{p\gamma}(\varepsilon_r)\, \kappa_{p\gamma}(\varepsilon_r)
\right\rangle,
\label{eq:tpgamma_def_1}
\end{equation}
where $\varepsilon$ is the photon energy in the plasma frame, $\varepsilon_r$ is the corresponding energy in the proton rest frame, $\sigma_{p\gamma}$ is the photomeson production cross section, and $\kappa_{p\gamma}$ is the inelasticity (the fraction of proton energy lost per interaction). The brackets denote an average over the pitch angle between the proton and photon momenta. For an isotropic photon field, the energy in the rest frame is $\varepsilon_r = \gamma_p \varepsilon (1 - \cos\theta)$, and the angular average can be performed explicitly, yielding:
\begin{equation}
t_{p\gamma}^{-1}(E_p)
=
\frac{c}{2\gamma_p^2}
\int_{\varepsilon_{\rm th}}^{\infty}
d\varepsilon_r\,
\sigma_{p\gamma}(\varepsilon_r)\,
\kappa_{p\gamma}(\varepsilon_r)\,
\varepsilon_r P(\varepsilon_r),
\label{eq:tpgamma_general}
\end{equation}
where
\begin{equation}
P(\varepsilon_r) = \int_{\varepsilon_r/(2\gamma_p)}^{\infty}
d\varepsilon\,
\varepsilon^{-2}\frac{d n_\gamma(\varepsilon)}{d \varepsilon}, 
\end{equation}
and $\varepsilon_{\rm th} \approx 145$ MeV is the threshold photon energy for pion production in the proton rest frame. Under the blackbody approximation, the photon spectrum follows a Planck distribution at temperature $T$:
\begin{equation}
\frac{dn_\gamma(\varepsilon)}{d\varepsilon}
=
\frac{8\pi}{h^3 c^3}\,
\frac{\varepsilon^2}{\exp(\varepsilon/k_{\rm B} T) - 1},
\label{eq:photon_bb}
\end{equation}
where $h$ is Planck's constant and $k_{\rm B}$ is the Boltzmann constant. Equation (\ref{eq:tpgamma_general}) then becomes:
\begin{align}
t_{p\gamma}^{-1}(E_p)
&=
\frac{4 \pi k_{\rm B} T}{\gamma_p^2 h^3 c^2} \times \nonumber \\
&\int_{\varepsilon_{\rm th}}^{\infty}
d\varepsilon_r\,
\sigma_{p\gamma}(\varepsilon_r)\,
\kappa_{p\gamma}(\varepsilon_r)\,
\varepsilon_r \,\,
\ln\left[
\frac{1}{1 - \exp\left(-\frac{\varepsilon_r}{2\gamma_p k_{\rm B} T}\right)}
\right].
\end{align}

This integral can be evaluated numerically without further approximations using tabulated cross sections $\sigma_{p\gamma}(\varepsilon_r)$ \citep{Morejon_2019} and inelasticities $\kappa_{p\gamma}(\varepsilon_r)$. The resulting cooling time is then expressed as a function of the neutrino energy $E_\nu \approx 0.05 E_p$.

\subsubsection{Proton--proton cooling time}

In addition to photomeson losses, protons may cool via inelastic proton--proton collisions in the dense medium. The corresponding cooling time is
\begin{equation}
t_{\rm {pp}}^{-1}(E_{\rm p})
=
c\,n_p\,
\sigma_{\rm pp}(E_{\rm p})\,
\kappa_{\rm pp}(E_{\rm p}),
\label{eq:tpp_def}
\end{equation}
where $n_p$ is the number density of target protons, $\sigma_{pp}(E_p)$ is the inelastic $pp$ cross section, and $\kappa_{pp}(E_p)$ is the inelasticity. The medium considered here has a proton number density $n_p \simeq 10^{17}$~cm$^{-3}$ and a mass density $\rho \simeq 10^6$~g~cm$^{-3}$. These values are not mutually constrained by $\rho = n_p m_p$ because most of the mass is carried by helium nuclei and degenerate electrons; the quoted $n_p$ represents the density of protons effectively available as hadronic targets. For the cross section, we adopt the fitting formulae from \citet{Kelner_ppsigma}:
\begin{equation}
    \sigma_{pp} = 34.3 + 1.88L + 0.25L^2,
\end{equation}
where $L = \ln(E_p/1\text{ TeV})$. The inelasticity is typically $\kappa_{pp} \sim 0.5$.

A similar cooling timescale can be defined for pions:
\begin{equation}
t_{\pi, \text{had}} = \frac{1}{c \sigma_{\pi p} n_p \kappa_{\pi p}},
\label{eq:10}
\end{equation}
where \(\sigma_{\pi p} \approx 5 \times 10^{-26}\)~cm$^2$ is the pion--proton scattering cross section and $\kappa_{\pi p} \approx 0.8$. Since we are evaluating the emission region, $n_p$ differs from the value in the remnant core. However, as the external layers were not involved in the hydrogen confinement, they likely preserve a low hydrogen density, though expansion following the core collision may further dilute the target material. To remain conservative, we use the same quantity $n_p$ as in the core; this overestimates $t_{\pi \text{had}}^{-1}$ and potentially leads to a stronger suppression factor than actually exists.

\subsubsection{Proton--alpha cooling time}

In the degenerate helium core environment, the number density of $^{4}\text{He}$ nuclei ($n_\alpha \sim 10^{29}$~cm$^{-3}$) is significantly higher than the density of free protons ($n_p \approx 10^{17}$~cm$^{-3}$). Consequently, inelastic collisions of relativistic protons with helium nuclei ($p\alpha$) must be considered. The inverse cooling time is:
\begin{equation}\label{eq:tpalpha_inv}
t_{p\alpha}^{-1}(E_p) = c n_\alpha \sigma_{p\alpha}(E_p) \kappa_{p\alpha}(E_p).
\end{equation}
At center-of-mass energies above $\sim 10$~GeV, the cross section can be described as $3.15 \times \sigma_{pp}(E_p)$ \citep{Orusa_2026}, and it holds that $\kappa_{p\alpha} \approx \kappa_{pp}$.

\subsubsection{Synchrotron cooling time of protons and pions}

Protons and pions lose energy via synchrotron radiation in the strong magnetic field. The synchrotron cooling times for protons and pions in the Thomson regime are:
\begin{equation}
t_{\rm syn} = \frac{6\pi m_p^4 c^3}{\sigma_{\rm T} m_e^2 B^2 E_p}, \quad t_{\pi,\text{syn}} = \frac{6\pi m_\pi^4 c^3}{\sigma_T m_e^2 B^2_{\rm r} E_\pi},
\label{eq:tsyn_p}
\end{equation}
where $\sigma_{\rm T} \approx 6.65 \times 10^{-25}$~cm$^2$ is the Thomson cross section. Here, $B_r$ represents the reconnection magnetic field. For our scenario, we adopt a reconnection rate of $R \sim 0.1/\sqrt{\beta} \approx 0.001$, where $\beta = P/(B^2/8\pi)$\citep{liu2024ohmslawreconnectionrate}.

\subsubsection{Proton inverse-Compton cooling time}

Protons also lose energy through inverse-Compton (IC) scattering off the thermal photon field. With $r_0 = e^2/m_e c^2 \approx 2.82 \times 10^{-13}$~cm and $\beta_p = \sqrt{1-\gamma^{-2}_p}$, the IC cooling time is:
\begin{equation}
t_{\text{IC}}^{-1} = \frac{c}{2\gamma_p^2} \left( \frac{m_e^2}{m_p^2} \right) \pi r_0^2 m_p^2 c^4 \int_0^\infty d\varepsilon \, \varepsilon^{-2} \frac{dn}{d\varepsilon} \frac{F(\varepsilon, \gamma_p)}{\beta_p(\gamma_p - 1)},
\end{equation}
where the function $F(\varepsilon, \gamma)$ accounts for the cross-section behavior across the Thomson and Klein-Nishina regimes using the auxiliary functions $f_1(z)$ and $f_2(z)$ defined via polylogarithms $\operatorname{Li}_2(z)$.

\subsubsection{Bethe--Heitler pair production cooling time}

Bethe--Heitler pair production, $p\gamma \rightarrow p e^+ e^-$, is treated analogously to photomeson production. The corresponding cooling time $t_{\rm BH}(E_p)$ uses the cross section:
\begin{equation}
   \sigma_{\rm BH} = \frac{28}{9}\alpha r^2_0 \ln\left[\frac{2 \epsilon E_p}{m_p m_e c^4}-\frac{106}{9} \right], 
\end{equation}
where $\alpha \approx 1/137$ is the fine structure constant and the threshold in the proton rest frame is $\approx 1.0226$~MeV.

\subsubsection{Pion decay cooling time}

The pion lifetime in the laboratory frame, where \(\tau_\pi \approx 2.6 \times 10^{-8}\)~s is the proper lifetime of \(\pi^\pm\) and \(\gamma_\pi\) is the Lorentz factor, is given by:
\begin{equation}
    t_{\rm \pi, dec} = \gamma_\pi \tau_\pi.
\end{equation}

\end{document}